\documentclass[aps,prb,preprint,amsmath,amssymb,superscriptaddress,showpacs,floatfix]{revtex4-1}

\usepackage{graphicx}
\usepackage{dcolumn}
\usepackage{bm}
\usepackage[colorlinks, allcolors=blue]{hyperref}
\usepackage[usenames, dvipsnames]{color}
\usepackage{soul}

\begin{document}

\title{Magnetocaloric effect in molecular spin clusters and their assemblies: Exact and Monte Carlo studies using exact cluster eigenstates}
\author{Sumit Haldar}
\email{sumithaldar@iisc.ac.in}
\affiliation{Solid State and Structural Chemistry Unit, Indian Institute of Science, Bengaluru - 560012, India.}
\author{S. Ramasesha}
\email{ramasesh@iisc.ac.in}
\affiliation{Solid State and Structural Chemistry Unit, Indian Institute of Science, Bengaluru - 560012, India.}


\begin{abstract}
Frustrated magnetic molecules are promising alternatives to refrigerant materials for low temperature magnetic refrigeration. We investigate the magnetocaloric effect (MCE) in un-frustrated and frustrated spin clusters formed from spin chains of six sites, with site spins $s=1$, $3/2$ and $2$ possessing site diagonal anisotropies and anisotropic exchange interactions, using exact diagonalization method. We also study MCE in spin clusters, on a chain, a 2-D square lattice and a 3-D cubic lattice with spin-dipolar interactions by a Monte Carlo method in spin-1 systems which uses exact eigenstates of a cluster. The magnetocaloric effect is closely related to the magnetic Gr\"uneisen parameter $\Gamma_H$. In this paper, we compute the magnetic Gr\"uneisen parameter $\Gamma_H$, and study its dependence on exchange anisotropy and spin-dipolar interaction. With increase of exchange anisotropy, the maxima in $\Gamma_H$ shifts to higher magnetic fields and becomes a sharp singularity. The singularities in $\Gamma_H$ correlate with cusps in the entropy as a function of magnetic field strength, and with crossover in the magnetization in the ground state in isolated clusters. The first maximum in $\Gamma_H$ shifts to lower fields as we increase spin-dipolar interaction. The first maximum in $\Gamma_H$ also shifts to lower magnetic field strength as the magnitude of the site spin increases. We show the dependence of $\Gamma_H$ on the dimensionality of the lattice for a fixed lattice constant.
\end{abstract}

\maketitle

\section{\label{sec:introduction}Introduction}
Magnetocaloric effect (MCE) has been extensively used in refrigeration by the adiabatic demagnetization (AD) method to obtain temperatures as low as a few micro Kelvin \cite{Launasmaa1974, Pecharsky1999, Tegus2002, Tishin2003}. In the AD method, a magnetic field is applied isothermally to a magnetic material and the field is removed adiabatically resulting in the cooling of the magnetic substance. The magnetic materials used in AD are mostly paramagnetic inorganic salts or oxides of rare earth elements because of the large magnetic moment associated with them \citep{Giauque1933, McMichael1992}. In recent years, there has been considerable interest in molecular magnets, in the quest for efficient MCE materials \cite{Envangelisti2005, Poddar2007, Ma2008, Affronte2004, Torres2003, Zhang2001, Evangelisti2010, Sessoli2012, Spickin2001}. While MCE was discovered more than a century ago \citep{Warburg1881}, the molecular magnets have a more recent history, of less than fifty years. Molecular magnets are usually polynuclear inorganic complexes of transition and/or rare earth metals \cite{Khan1993, Gatteschi2007, Wang2011}. The metal centers are magnetic and have exchange interactions among them. The resulting spin of the molecule is large and hence the magnetic entropy associated with these systems is also expected to be large. Thus, one would expect a large increase in magnetic entropy when the magnetizing field is switched off adiabatically. This should translate to a large MCE, since to compensate the increase in magnetic entropy, the lattice entropy will decrease, resulting in the cooling of the sample. Indeed there have been many studies on the well known single molecule magnets (SMMs) $Mn_{12}$ and $Fe_{8}$ \cite{Torres2003, Zhang2001, Envangelisti2005, Spickin2001}, in which large MCE has been observed but the effect diminishes at liquid Helium temperatures due to thermal blocking of the magnetization. However, there is also recent report that says SMMs are not good magneto caloric materials \cite{Beckmann2019}. There have been studies on MCE of one-dimensional antiferromagnets and high spin cycles \cite{Zhitomirsky2004, Baniodeh2018}.

The SMMs are characterized by frustrated exchange interactions between anisotropic magnetic centers \cite{Raghu2003}. There have been many theoretical studies in SMMs and single chain magnets (SCMs) and the role of on-site anisotropies and exchange anisotropies in determining the overall magnetic anisotropy of the SMMs and SCMs \cite{Haldar2017, Haldar2018}. Frustration in the exchange interaction leads to degeneracy close to the ground state and thus large variations in the magnetic entropy when the material is demagnetized \cite{Zhitomirsky2003, Schnack2007, Garlatti2012, Sharples2014}. Indeed a large MCE in comparison with previous known values, has been found in a $Fe_{14}$ molecular cluster which is characterized by large spin ground state, small magnetic anisotropy, and high density of states close to the ground state \cite{Evangelisti2005}. Later on, Zheng $\textit{et al.}$ synthesized a Wells-Dawson type \{$Ni_6Gd_6P_6$\} cage \cite{Zheng2011}. From the magnetism data, they have reported that the system of spins has alternating ferromagnetic and antiferromagnetic nearest neighbour interaction and next-nearest neighbour ferromagnetic interaction present between \{$Ni_3$\} triangle and weak interactions between lanthanide ions and $Ni^{2+}$ ions. However, we have not seen any systematic theoretical study of the MCE in SMMs or SCMs in the literature. In this paper, we study MCE in a class of spin chains which are characterized by frustrated next nearest neighbor exchange interactions, on-site magnetic anisotropy and exchange anisotropy for nearest neighbor interactions. We employ the exact diagonalization method to study MCE in small spin chains of spin $s=1$, $3/2$ and $2$ with site diagonal anisotropy. We also study an assembly of these spin chains with spin-dipolar interactions using the Monte Carlo method. Specifically, we deal with two kinds of spin chains (i) alternating ferromagnetic (F) and antiferromagnetic (AF) exchange interactions between nearest neighbors and (ii) the above model with an additional next nearest neighbor ferromagnetic interactions. Experimentally, such systems have been realized in $Ni_6Gd_6P_6$, $(CH_{3})CHNH_{3}CuCl_{3}$, $CuNb_{2}O_{6}$ and $M[(4, 4^{'}-dimethylbipyridine)(N_{3})_{2}]_{n}[M=Cu(II), Mn(II), Ni(II) and Fe(II)]$ \cite{Zheng2011, Kodama1999, Manaka1997, Stone2007, Shen2000}. In the first model there is no frustration while in the second model frustration is built in due to the next nearest neighbor ferromagnetic interactions. In the next section we introduce the model spin Hamiltonians and discuss numerical method for obtaining the MCE coefficient characterized by the magnetic Gr\"uneisen parameter $\Gamma_H$. In section 3 we discuss the dependence of $\Gamma_H$ on frustration and magnetic anisotropy, from both exact diagonalization and Monte Carlo studies. In the last section we summarize our results.

\section{\label{sec:Methodology}Model Hamiltonian and Methodology}
\begin{figure}[hbt!]
    \includegraphics[width=12cm]{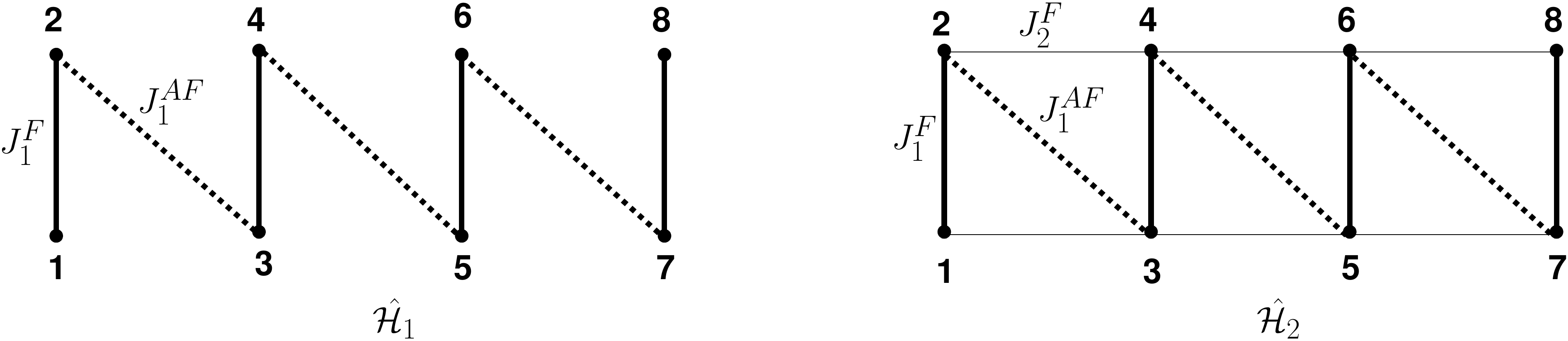}  
      \caption{\label{fig:modelfig}Schematic diagrams of the exchange interactions between spins in the two models described by Hamiltonians $\hat{\mathcal{H}_{1}}$ and $\hat{\mathcal{H}_{2}}$ shown for an open chain of eight sites. The nearest neighbor exchange interactions are anisotropic and have the same magnitude of anisotropy irrespective of the nature (ferro/antiferro) of the exchange. Second neighbor interactions are always isotropic.}
\end{figure}

The Hamiltonian of the two models we have studied are given by
\begin{eqnarray}
\label{eqn:NonFrustratedHamil}
\hat{\mathcal{H}_1} &=&-J_1 \Biggl[\sum_{k=1}^{N/2} (\hat{s}_{2k-1}^z \hat{s}_{2k}^z - \hat{s}_{2k}^z \hat{s}_{2k+1}^z)+(1-\epsilon) \biggl\{(\hat{s}_{2k-1}^x \hat{s}_{2k}^x - \hat{s}_{2k}^x \hat{s}_{2k+1}^x) \nonumber \\ 
 & & \qquad +  (\hat{s}_{2k-1}^y \hat{s}_{2k}^y-\hat{s}_{2k}^y \hat{s}_{2k+1}^y)\biggr\} \Biggr] - \sum_{k=1}^{N} g\mu_B \hat{s}_k^z h^z + d\sum_{k=1}^{N} \hat{s}_{k}^{z^{2}},
\end{eqnarray}
and
\begin{eqnarray}
\label{eqn:FrustratedHamil}
\hat{\mathcal{H}_{2}}=\hat{\mathcal{H}_{1}} - J_2 \sum_{k=1}^{N-2} \vec{s}_{k} \cdot \vec{s}_{k+2},
\end{eqnarray}
where the summations imply chosen boundary conditions, $h^z$ is the magnetic field along the $z$-axis. The nearest neighbor (nn) exchange is alternating ferromagnetic and antiferromagnetic, with the same magnitude $|J_1|$, $\epsilon$ is the deviation of the nn exchange from isotropy and is taken to be the same for both ferro and antiferro magnetic exchanges and $\hat{\mathcal{H}_{1}}$ reduces to an Ising model for $\epsilon=1$. The exchange interactions are shown schematically in Fig. \ref{fig:modelfig}.
The site anisotropy of the spins is assumed to be axial with the magnitude $d$ for all sites. We have taken $d$ to be negative so that the resulting spin cluster is magnetic and it also defines the easy axis of magnetization. The nnn interaction is taken to be ferromagnetic. The off-diagonal site anisotropy is usually very much smaller than $d$ and plays an important role in studying the dynamics of magnetization relaxation.

Both $\hat{\mathcal{H}_{1}}$ and $\hat{\mathcal{H}_{2}}$ conserve the z-component of total spin and hence the Hamiltonian matrix is block diagonal in $M_s$. We solve for all the eigenvalues and eigenvectors in all the $M_s$ sectors as we are interested in computing thermodynamic properties. The quantity we compute is the magnetic Gr\"uneisen parameter $\Gamma_H$ given by
\begin{eqnarray}
\label{eqn:Gruneisenratio}
\Gamma_{H}=\frac{1}{T}\left(\frac{\partial T}{\partial H}\right)_{S}=-\frac{1}{C_H}\left(\frac{\partial M}{\partial T}\right)_H,
\end{eqnarray}
where the symbols have the usual meaning and the equality is obtained from Maxwell relation. A larger $\Gamma_H$ implies higher cooling efficiency in the AD process. The specific heat, $C_H$ and the derivative $(\frac{\partial M}{\partial T})_H$ can be obtained from the relations
\begin{eqnarray}
\label{eqn:specificheat}
C_{H}=\frac{(\langle E^{2} \rangle-\langle E \rangle^{2})}{T^{2}};
\end{eqnarray}

\begin{eqnarray}
\label{eqn:delMbydelT}
\left(\frac{\partial M}{\partial T}\right)_{H}= \frac{(\langle ME \rangle-\langle M \rangle \langle E \rangle)}{T^{2}}.
\end{eqnarray}
Thus, computation of $\Gamma_H$ is carried out from the full eigenvalue spectrum of the Hamiltonian for chain lengths $N=6$ and $8$ and site spins $s=1$ and for chain length $N=6$ for spins $3/2$ and $2$. The entropy of a single cluster is calculated explicitly from the partition function as 
\begin{eqnarray}
\label{eqn:entropy}
S=\frac{1}{TZ} \sum_n E_n e^{-E_n/T}+lnZ,
\end{eqnarray}
where $Z$ is the partition function.

We have also carried out MCE calculations on an assembly of the spin clusters. We have assumed classical spin dipolar interaction between clusters, whose energy is given by
\begin{eqnarray}
\label{eqn:dipolar}
E_{ij}^{dip}=\frac{\vec{M}_i \cdot \vec{M}_j}{r_{ij}^3} - 3 \frac{(\vec{M}_i \cdot \vec{r}_{ij})(\vec{M}_{j} \cdot \vec{r}_{ij})}{r_{ij}^5}.
\end{eqnarray}
In the Monte Carlo calculations, which uses the Metropolis algorithm, the states of the Markov chain comprises of all the eigenstates of all the clusters in the system, namely the set \{$i_k$\} where $i_k$ is the $i^{th}$ state of the $k^{th}$ molecule with z-component of the magnetization given  by $M_{i,k}$. We neglect the x and y component of the magnetization in treating the dipolar interactions.

\begin{figure}[hbt!]
    \includegraphics[width=6cm]{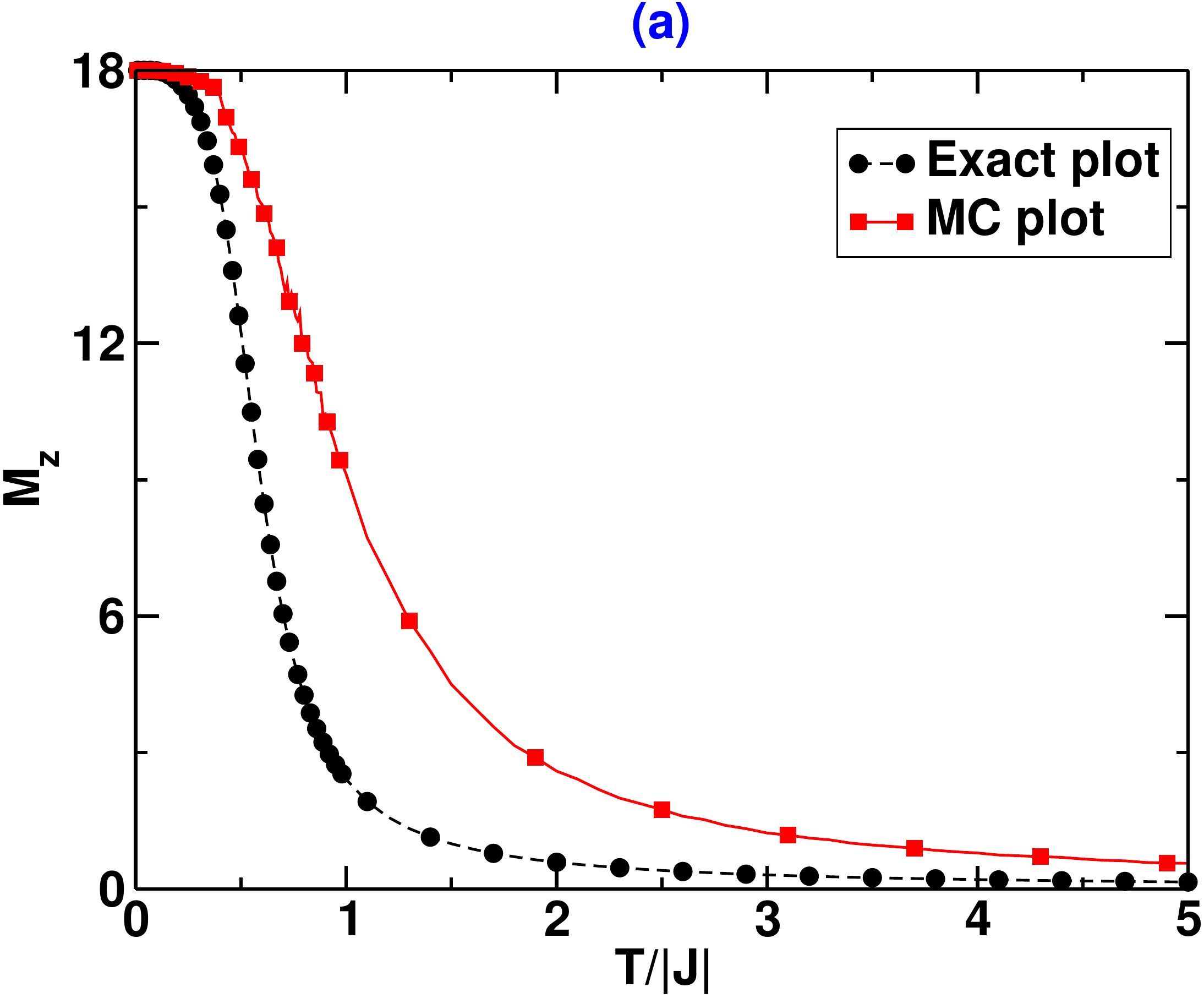} \hspace{0.3in}  \includegraphics[width=6cm]{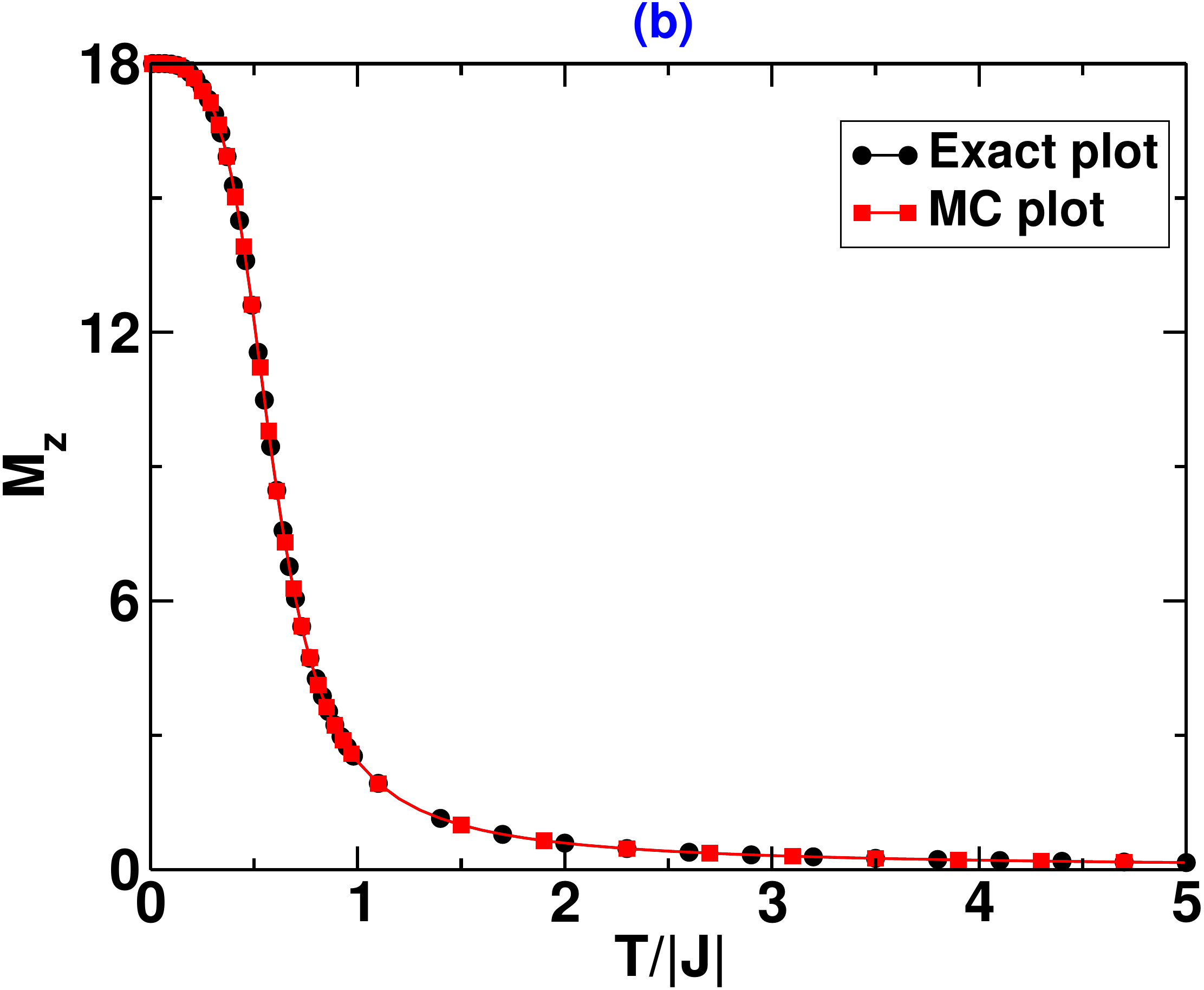} 
      \caption{\label{fig:exactvsMC}Dependence of magnetization ($M_z$) on temperature ($T/|J|$) for a system with three spin clusters of six sites with site spin $s=1$ and on-site anisotropy $|d/J|=0.1$ in the presence of magnetic field $g\mu_B H/|J|=0.05$, computed by two different methods namely by the Monte Carlo method and exact diagonalization. In Fig. (a), Monte Carlo calculations are done using single flip mechanism and in Fig. (b) employs multiple spin flip mechanism.}
\end{figure}

In the implementation of the MC algorithm, we choose a site $k$ at random and choose a state $i^{'}$ to which we wish to make a transition from the initial state $i$. The energy difference $\Delta E$ for this change is given by
\begin{eqnarray}
\Delta E=E_{i^{'}}-E_i + \sum_l \left[\frac{(M_{i^{'}k}-M_{i,k}) M_{j,l}}{r_{k,l}^3}-3\frac{(M_{i^{'}k}-M_{i,k}) M_{j,l} z_{k,l}^{2}}{r_{k,l}^5}\right]
\end{eqnarray}
If $\Delta E$ is -ve the state of the spin cluster at site $k$ is changed from $i$ to $i^{'}$ otherwise it is changed with a probability $\textrm \{ exp(-\frac{\Delta E}{T})$\}. This is distinct from the single spin flip mechanism as $M_{i^{'},k}$ does not necessarily differ from $M_{i,k}$ by unity. We found that restricting to single spin flip mechanism ($\Delta M=\pm1$) does not yield the correct thermodynamic properties even after a very large number of MC steps, in the case of two and three spin clusters for which exact thermodynamic calculation can be carried out. Hence, we have used a general mechanism in which the state at site $k$ can flip from any magnetization $M_{k,i}$ to any other magnetization $M_{k,i^{'}}$. In Fig. \ref{fig:exactvsMC},
we have compared exact results for a three spin-clusters with Monte Carlo calculations using single and multiple spin flip algorithms. We find that the multiple spin flip algorithm reproduces exact results while single spin flip mechanism fails to converge to exact values. We have also estimated errors in the case of the chains by using the thermal averaged data for averages obtained at intervals of 1 million steps from 491 million to 500 million MC steps and fitting these averages using a linear least squares fit. The estimated error bars are shown in Fig. \ref{fig:largesystemGamma}. All our simulations employ multiple spin flip algorithm and the number of MC steps for thermalization is 500 million and the thermal averaging is done over the next 500 million steps. In the case of a $5 \times 5 \times 5$ lattice, thermalization was carried out for 50 billion steps and averaging is done over further 50 billion steps to confirm the oscillations in the Gr\"uneisen parameters. We have studied a chain, a 2-D square lattice and a 3-D simple cubic lattice with up to 125 spin clusters. In the case of the chain, the anisotropy $d$ is along the chain axis, which is the z-axis. In the 2-D and 3-D cases the anisotropy is along one of the unit cell axes.

\section{\label{sec:Results}Results and Discussion}
We have carried out studies on the two models for different site spins, for single cluster as well as for 1, 2 and 3-D assemblies, each involving about a hundred spin clusters. We have computed the magnetic Gr\"uneisen parameter for single cluster as well as for the cluster assemblies.
\subsection{\label{sec:Nonfrustratedsystem}Non-frustrated Spin Chains (Model I)}
\begin{figure}[hbt!]
    \includegraphics[width=8cm]{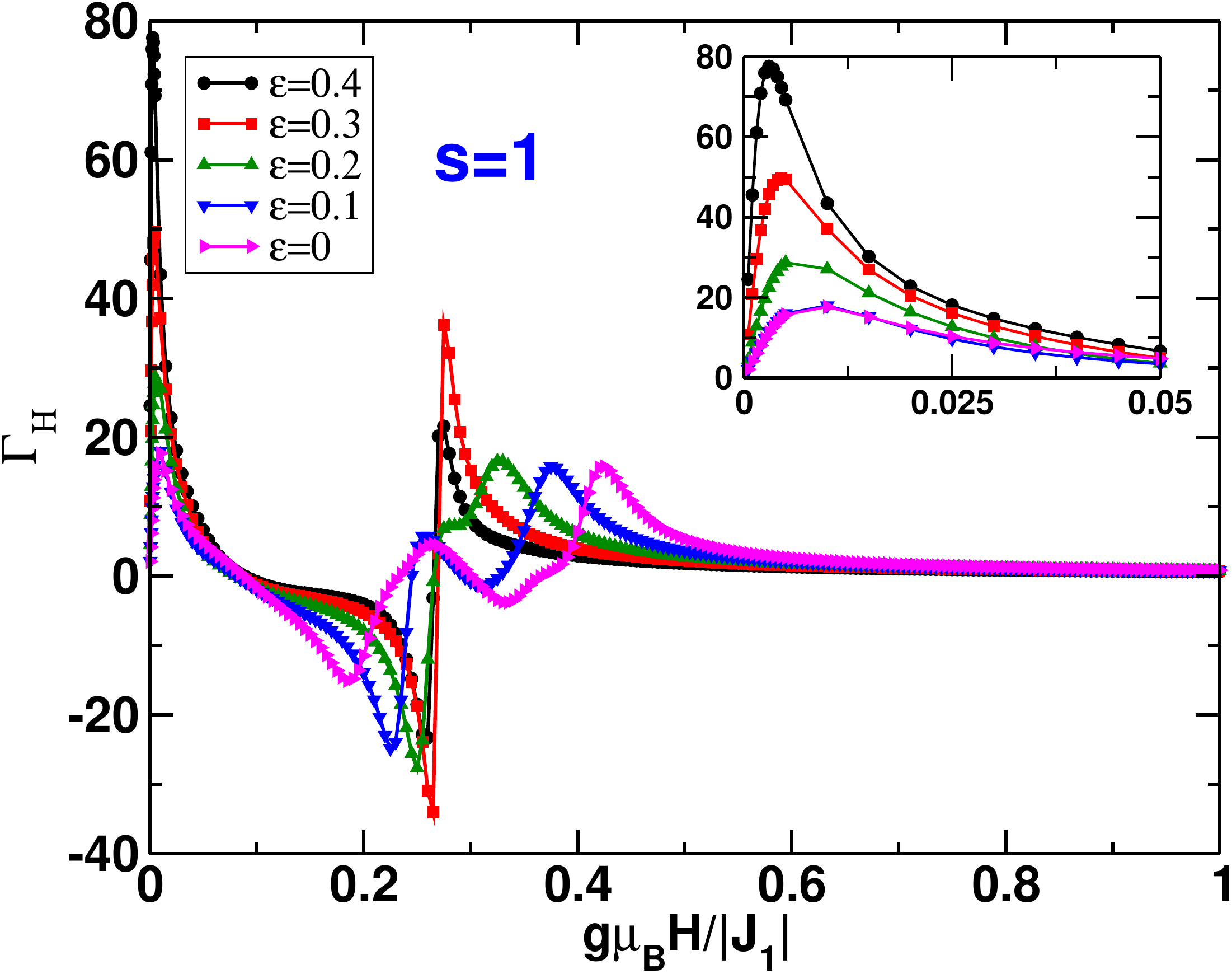} \\
     \includegraphics[width=8cm]{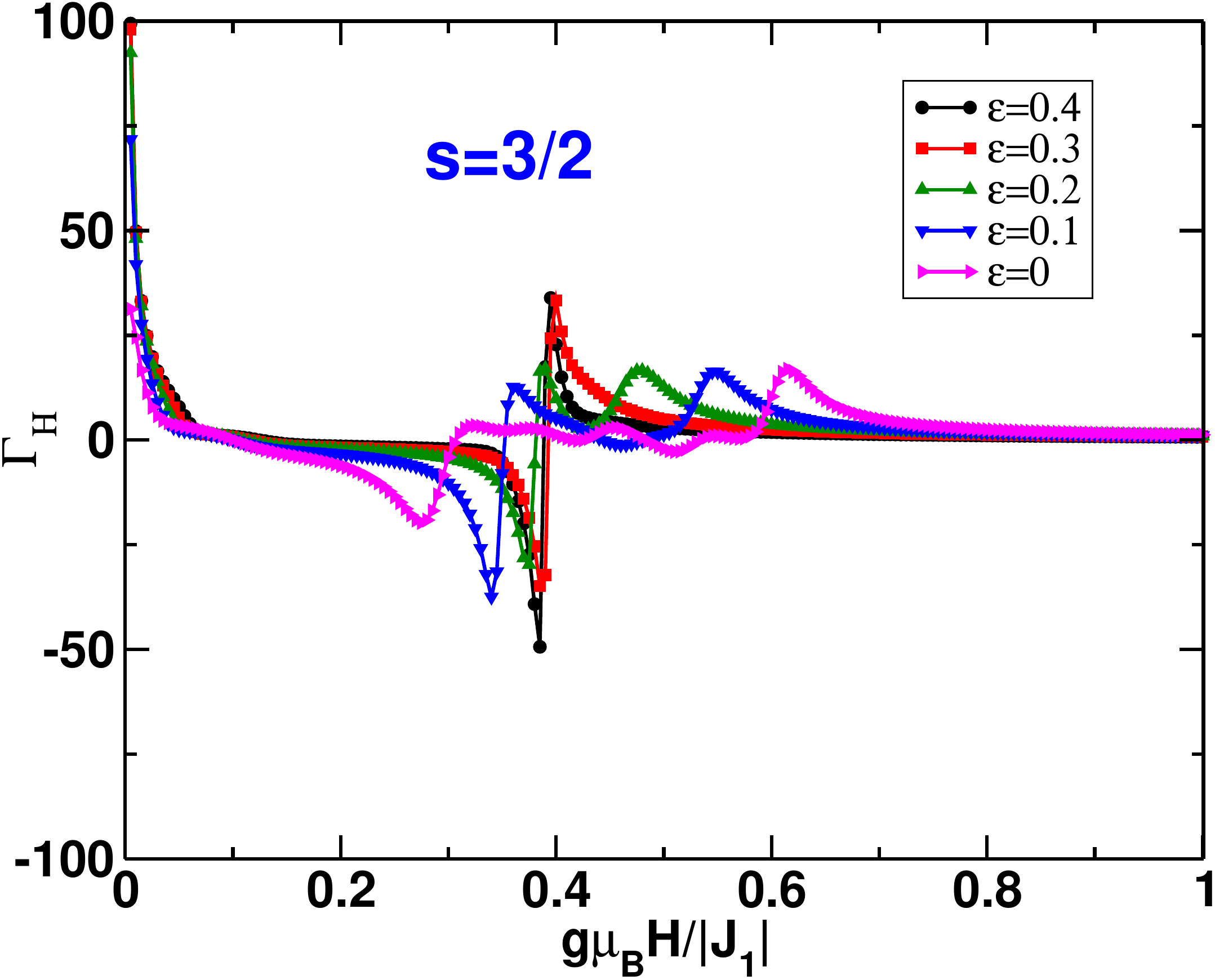} \\  
     \includegraphics[width=8cm]{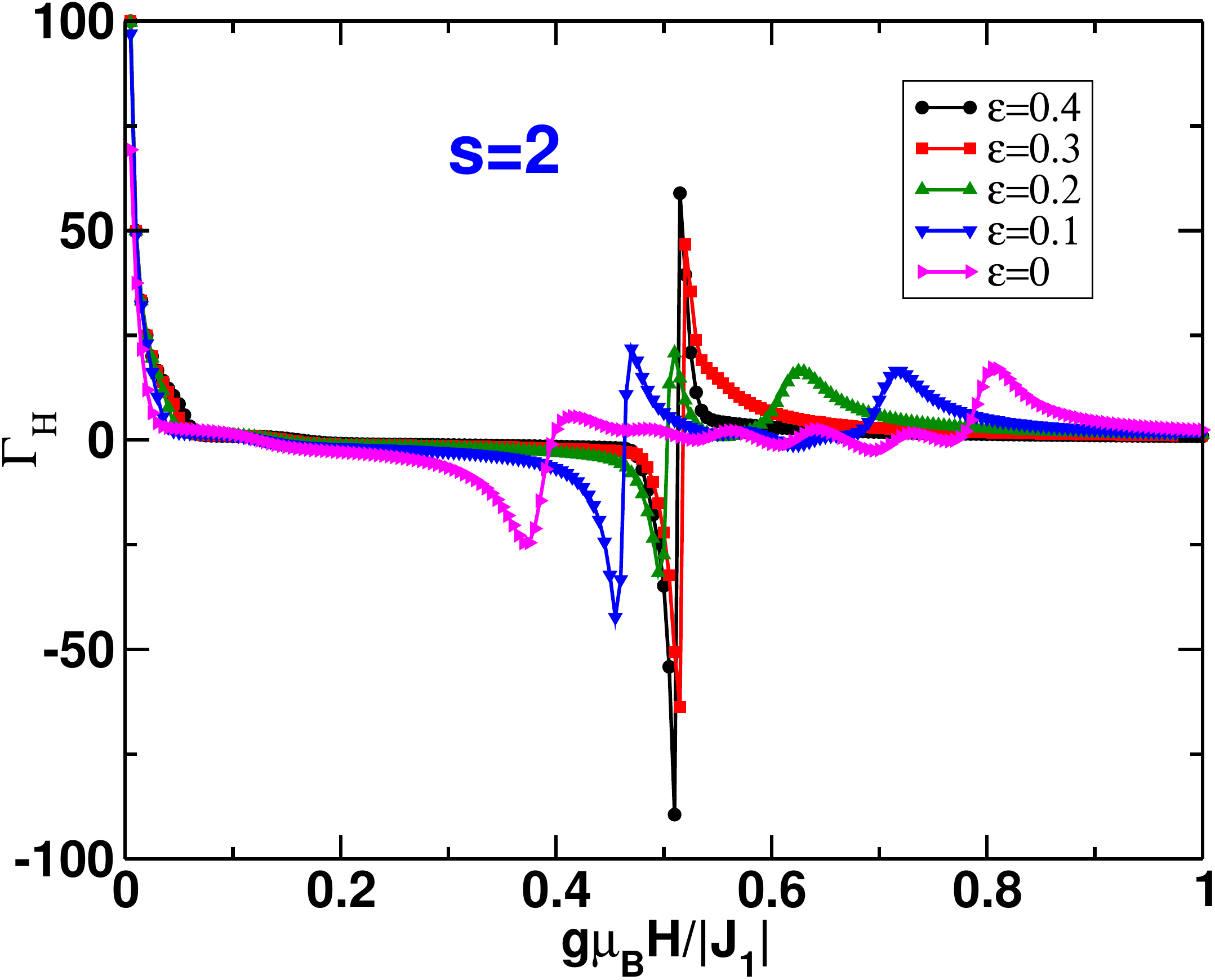} 
      \caption{\label{fig:GruneisenNonfrustrated}Variation of Gr\"uneisen parameter, $\Gamma_{H}$, with applied magnetic field ($g\mu_B H/|J_1|$) at temperature $k_BT/|J_1|=0.1$ for different values of exchange anisotropy $\epsilon$, in the presence of on-site anisotropy $|d/J_1|=0.1$ for spin chains with site spins $s=1$, $3/2$, and $2$ in chains of six sites. Inset in the top panel is for low magnetic field strengths.}
\end{figure}
The exact magnetic Gr\"uneisen parameter $\Gamma_{H}$ is shown in Fig. \ref{fig:GruneisenNonfrustrated} for clusters of six spins with site spins $s=1$, $s=3/2$ and $2$. We note that $\Gamma_H$ peaks in the low field limit in all the cases. The peak is higher for larger exchange anisotropy. These features are consistent with the fact that the system is magnetic in the ground state and the magnetization of the system increases with anisotropy in the exchange constant $J_1$ and the site spins. What is interesting is the behaviour  of $\Gamma_{H}$ in the higher field regime. For isotropic exchange, there are oscillations in $\Gamma_{H}$ at higher fields. As the exchange anisotropy is increased all these systems give rise to a single sharp minima followed by a maxima, arising from a level crossing. This is true for all site spins, although the magnetic field at which this strong criticality is observed increases with the value of the site spin.
\begin{figure}[hbt!]
    \includegraphics[width=8cm]{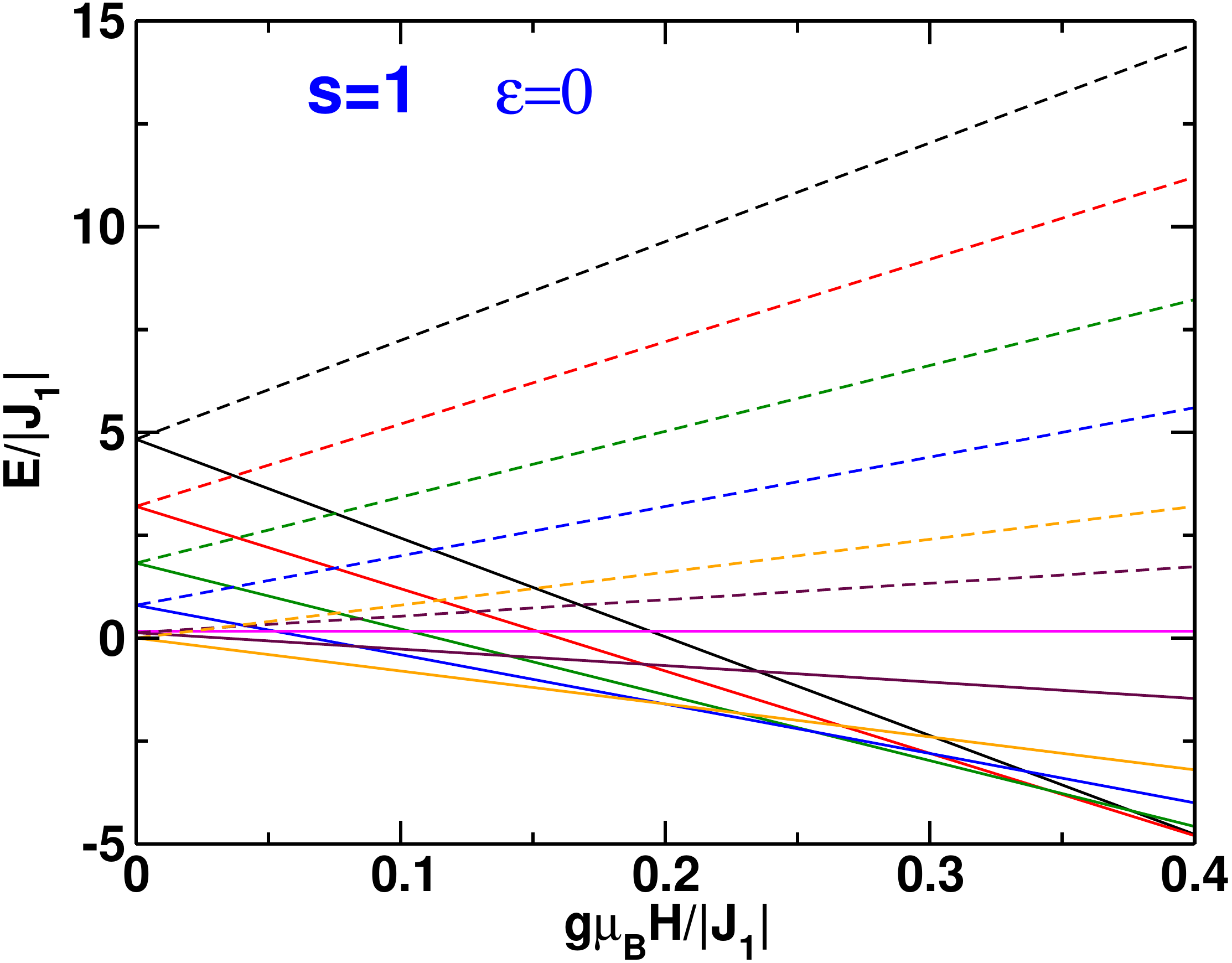}\\
     \includegraphics[width=8cm]{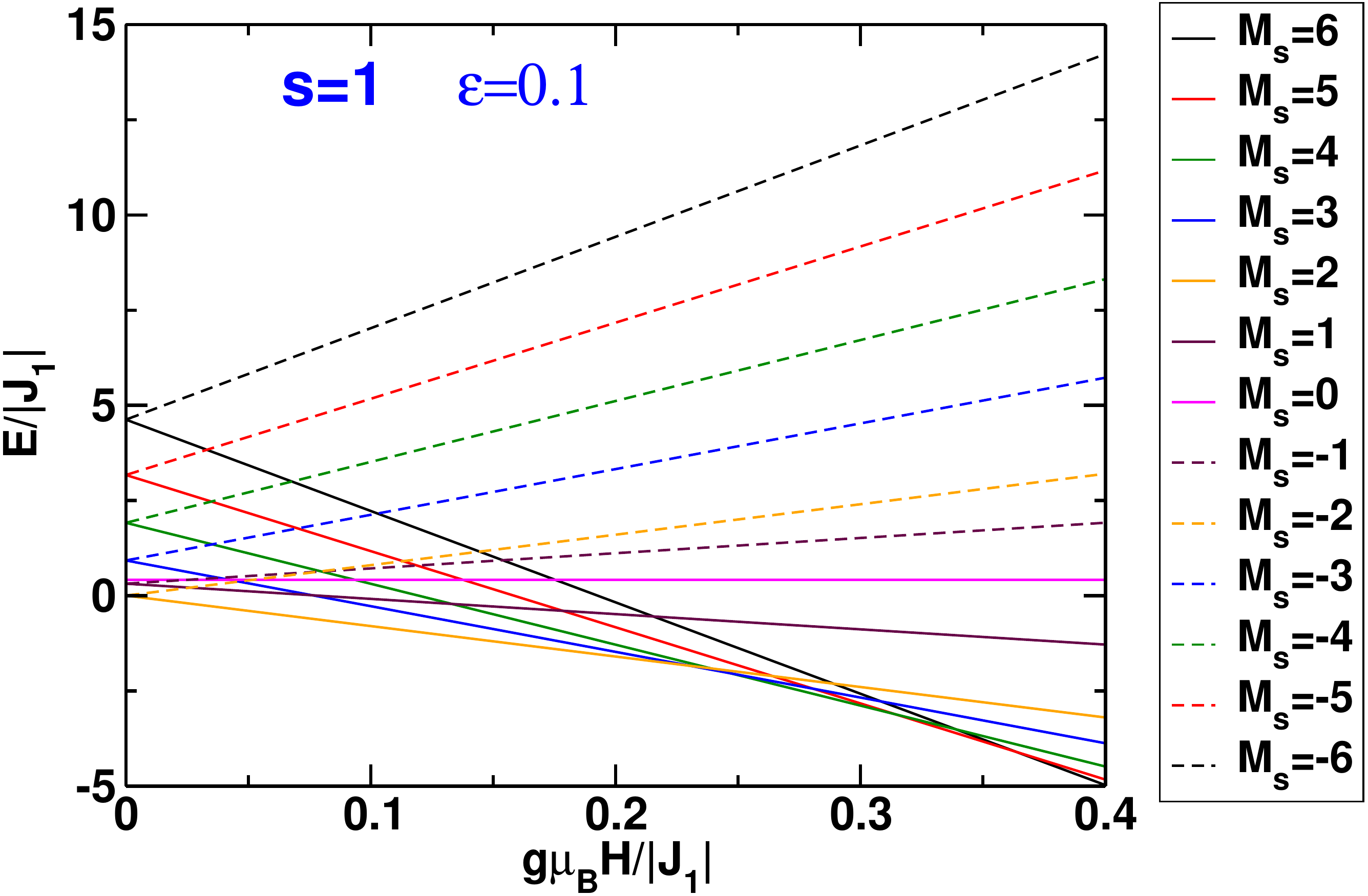} \\
     \includegraphics[width=8cm]{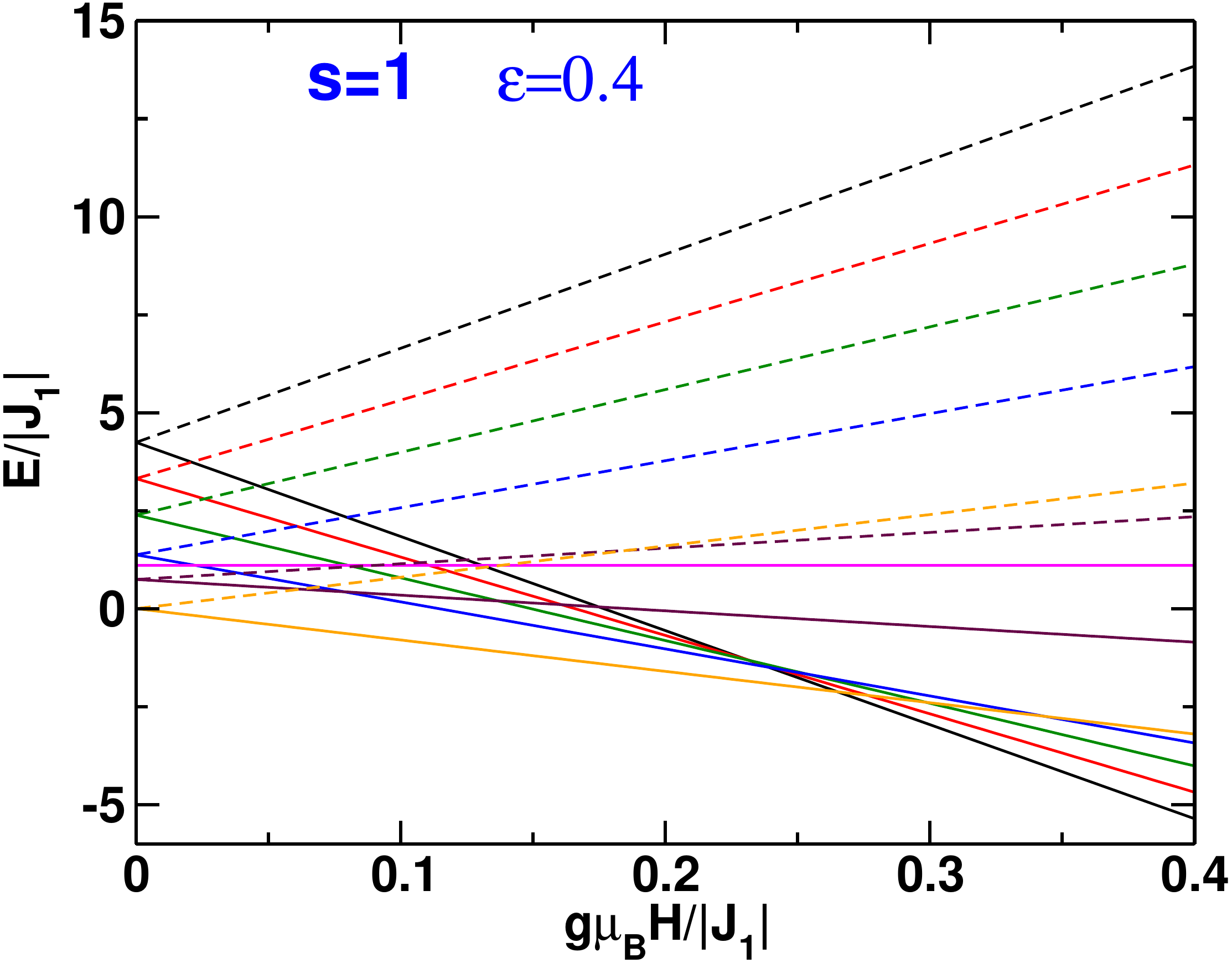} 
      \caption{\label{fig:zeemanplot}Field dependence of few low lying states above the ground state for the exchange anisotropies $\epsilon=0$, $0.1$ and $0.4$ at temperature $k_BT/|J_1|=0.1$, in the presence of on-site anisotropy $|d/J_1|=0.1$ for spin chains of six sites with site spin $s=1$. Solid and broken lines are for different $\pm M_s$ values which are color coded as shown in the side bar.}
\end{figure}

\begin{figure}[hbt!]
    \includegraphics[width=8cm]{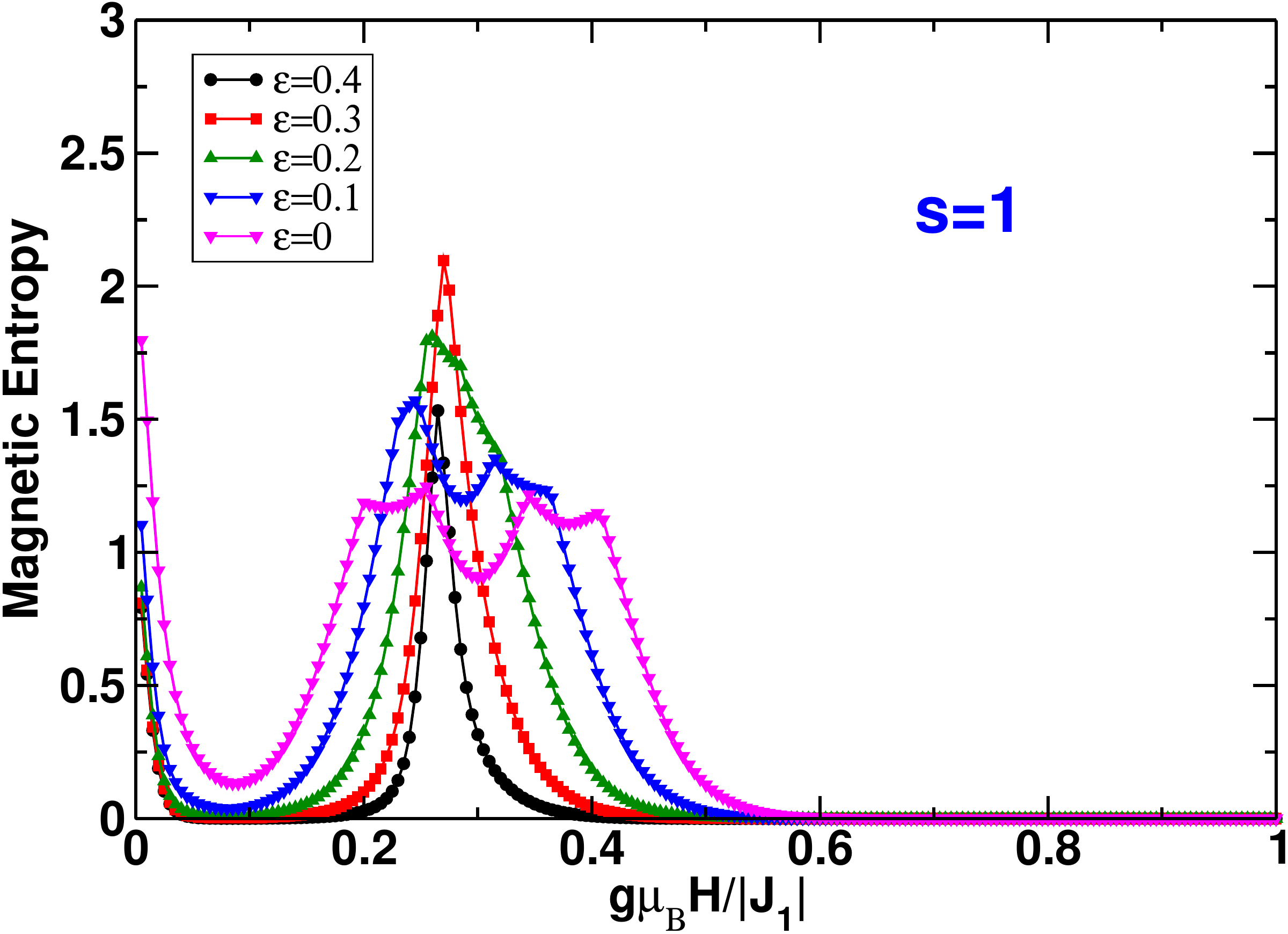} \\
     \includegraphics[width=8cm]{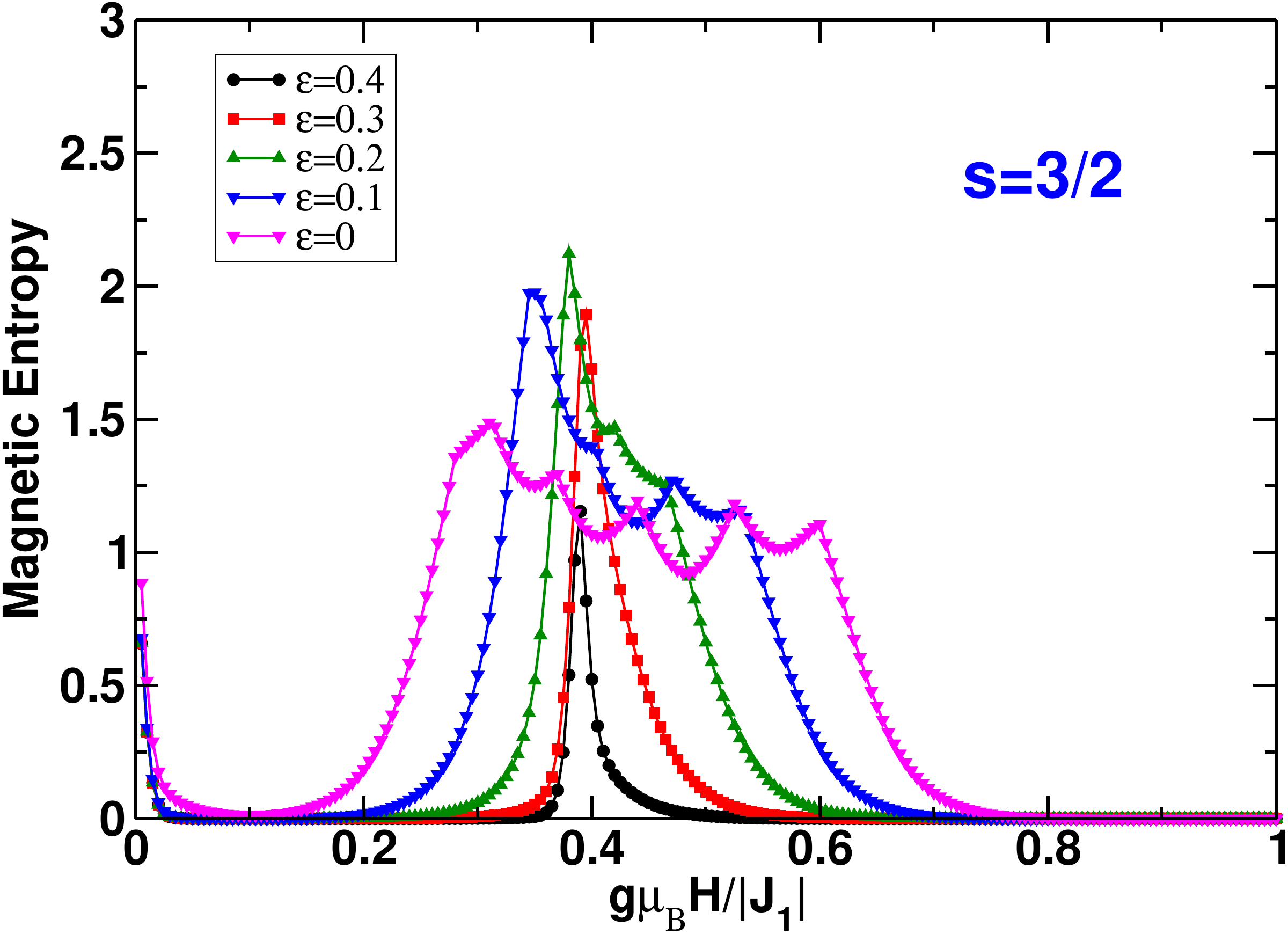}  \\
     \includegraphics[width=8cm]{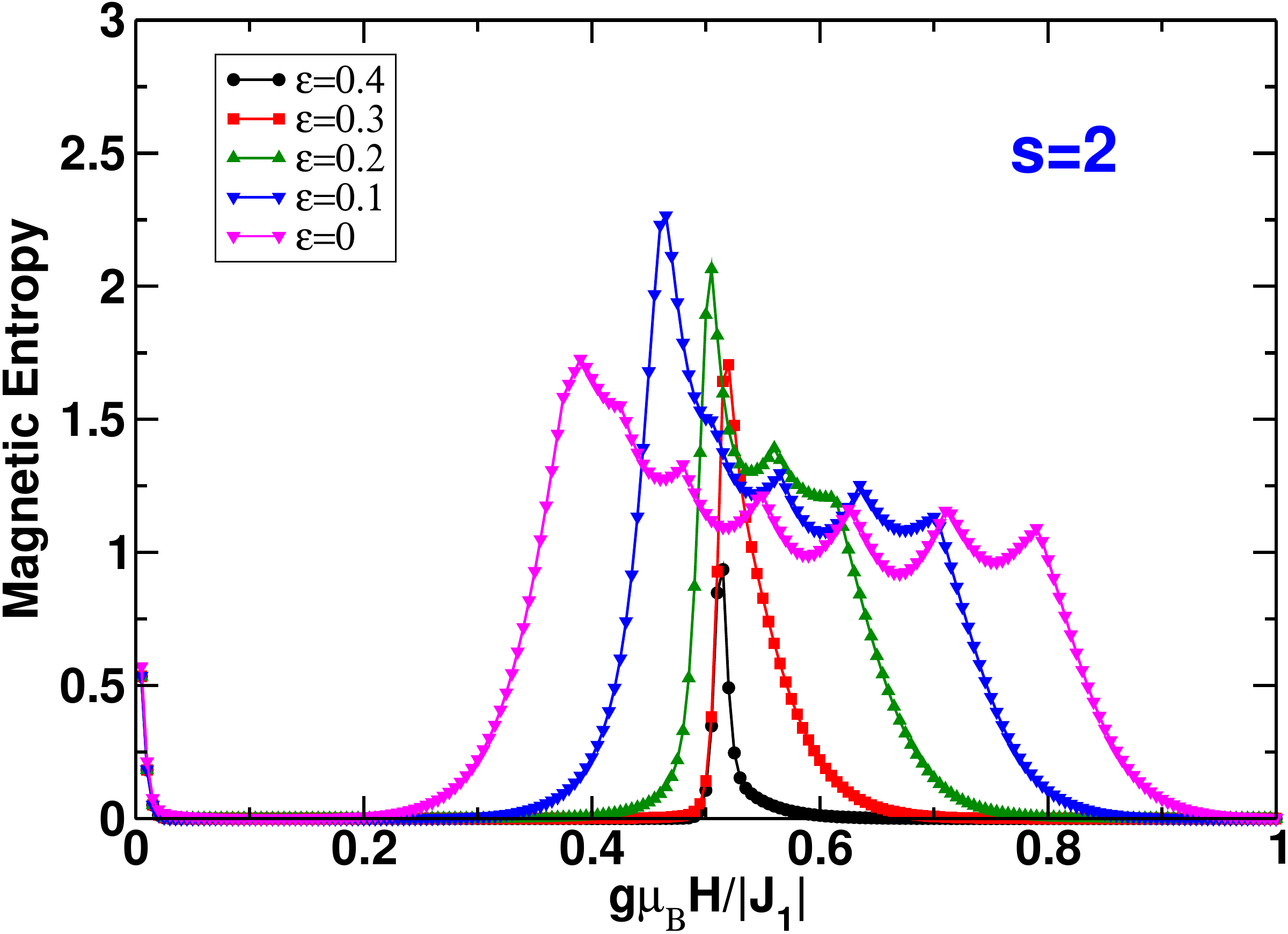} 
      \caption{\label{fig:entropy}Variation of magnetic entropy, with applied magnetic field ($g\mu_B H/|J_1|$) at temperature $k_BT/|J_1|=0.1$ for different exchange anisotropy values $\epsilon$, with on-site anisotropy $|d/J_1|=0.1$, for spin chains of six sites with site spins $s=1$, $3/2$, and $2$.}
\end{figure}

To understand this behaviour, we have examined the evolution of the low-lying energy levels as a function of the magnetic field. We note that for isotropic exchange, there are several energy level crossings of states with different magnetization, $M_s$ (Fig. \ref{fig:zeemanplot}). We find broad oscillations (for small exchange anisotropy) or sharp singularities (for large exchange anisotropies) at magnetic fields at which the ground state $M_s$ value changes. For low-anisotropy, the crossover in the ground states $M_s$ value occurs between $M_s=2$ and $M_s=3$, with $M_s=4$ closely following the crossover from $M_s=3$. The crossover in the ground state $M_s$ is from $M_s=2$ to $M_s=6$ and there are no other crossovers in $M_s$ after this.  These energy level crossovers also manifest as cusps in the magnetic entropy vs magnetic field (Fig. \ref{fig:entropy}) for all the different site spins.
 The sharp cusp in the magnetic entropy when the ground state $M_s$ value changes manifests as sharp singularities in the magnetic Gr\"uneisen constants.
 
We have also studied the behaviour of the magnetic Gr\"uneisen parameters, in an assembly of spin clusters with inter-chain spin dipolar interactions. We have carried out these studies on a chain of 100 spin clusters, a $10 \times 10$ square lattice and a $5\times 5\times 5$ simple cubic lattice with 125 spin clusters. The nearest neighbor spin dipolar interaction energy $E_{12}^d$ are chosen to be $1.2J_1$, $2.4J_1$ and $6J_1$ for a fully polarized spin 6 cluster (M=6) with magnetization oriented along chain axis for a pair of nearest neighbor clusters. This in turn scales with the nearest neighbor distance. The lattice constant of the system in all dimensions are taken to be unity. The Monte Carlo calculations have been carried out only on the spin-1 systems. 

\begin{figure}[hbt!]
    \includegraphics[width=13cm]{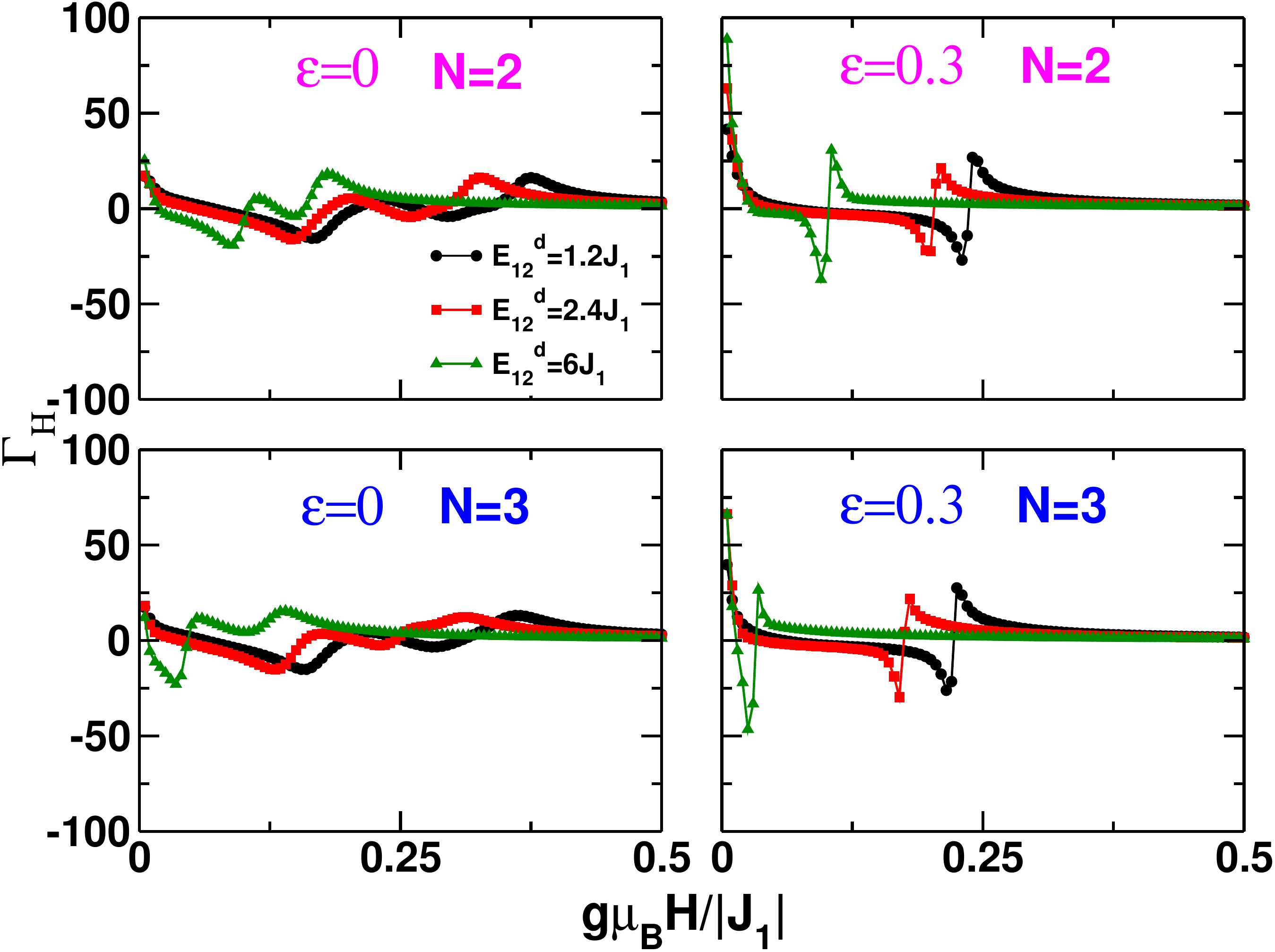}
      \caption{\label{fig:smallsystemGamma}Variation of Gr\"uneisen parameter, $\Gamma_H$, with applied magnetic field ($g\mu_B H/|J_1|$) for two small spin clusters $N=2$ and $3$ on a chain with different values of spin-dipolar interactions $E_{12}^d=1.2J_1$, $2.4J_1$ and $6J_1$ for exchange anisotropies $\epsilon=0$ and $0.3$ and on-site anisotropy $|d/J_1|=0.1$ for spin cluster with site spin $s=1$ at temperature $k_BT/|J_1|=0.1$.}
\end{figure}

\begin{figure}[hbt!]
    \includegraphics[width=13cm]{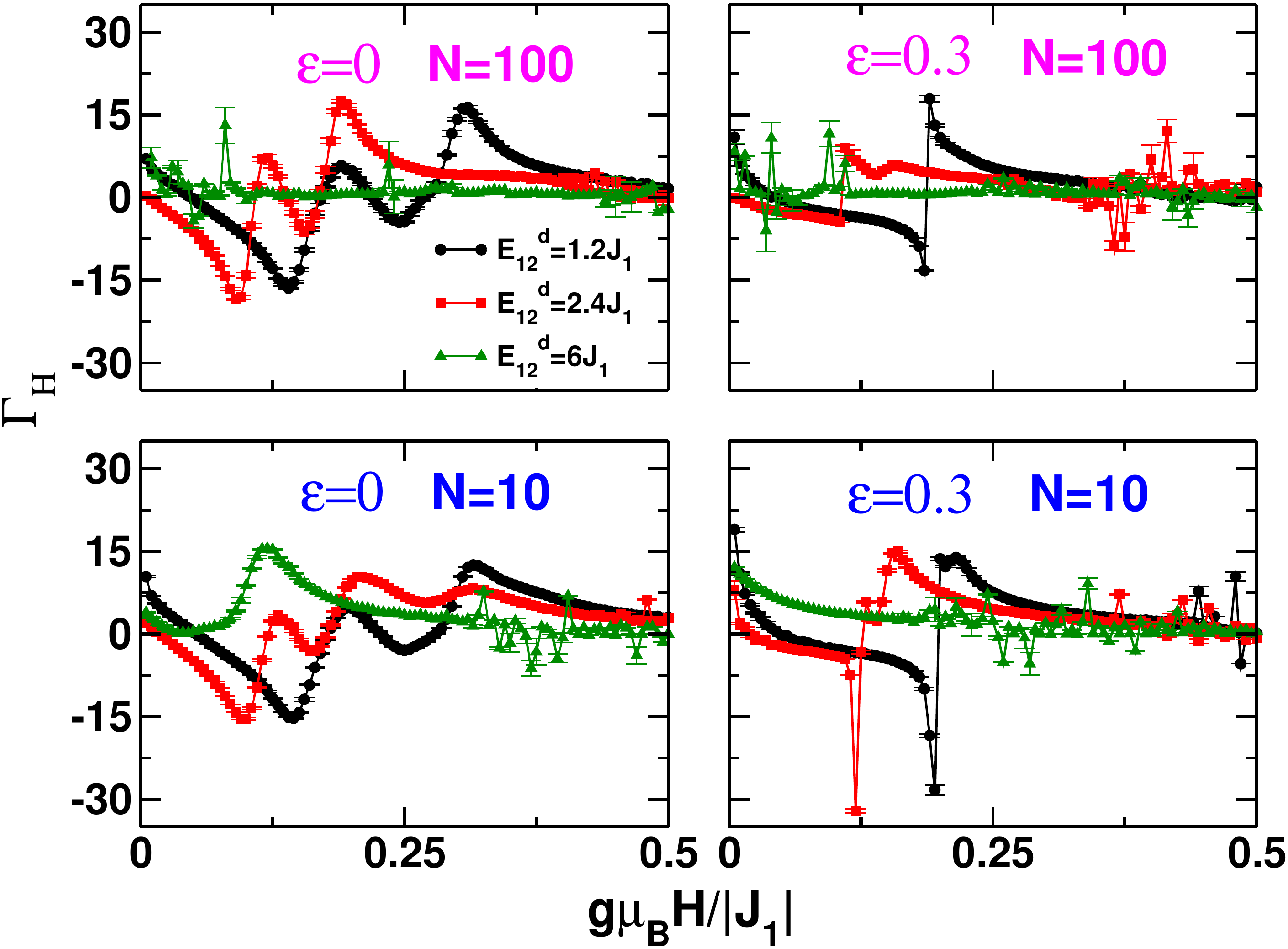}
      \caption{\label{fig:largesystemGamma}Variation of Gr\"uneisen parameter, $\Gamma_H$, with applied magnetic field ($g\mu_B H/|J_1|$) for two larger spin clusters with $N=100$ and $10$ on a chain with different values of spin-dipolar interactions $E_{12}^d=1.2J_1$, $2.4J_1$ and $6J_1$ for exchange anisotropies $\epsilon=0$ and $0.3$ and on-site anisotropy $|d/J_1|=0.1$ for spin cluster with site spin $s=1$ at temperature $k_BT/|J_1|=0.1$. Error bars for the estimation are also shown for each data point.}
\end{figure}

We obtained the Gr\"uneisen parameter $\Gamma_H$ for systems of $N=2$, $3$, $10$ and $100$ clusters for dipolar interactions strength $E_{12}^d$=$1.2J_1$, $2.4J_1$ and $6J_1$ and nearest neighbor exchange anisotropies $\epsilon=0.3$ and $0$. The system sizes $N=2$ and $3$ were chosen also as a test of the Monte Carlo algorithm, as for these sizes, exact calculations are feasible. In small system Fig. \ref{fig:smallsystemGamma}
with $N=2$ and $3$ singularities in $\Gamma_H$ shifts to lower fields as the strength of dipolar interactions increases. For systems sizes with $N=10$ and $100$ Fig. \ref{fig:largesystemGamma}
the low field singularity of $\Gamma_H$ nearly vanishes for large anisotropy and strong dipolar interaction. In the isotropic model as in the isolated cluster case there are several broad singularities of $\Gamma_H$ which become sharper with increasing anisotropy parameter $\epsilon$. However when the dipolar interaction becomes strong then these singularities are suppressed considerably in all cases.

\begin{figure}[hbt!]
    \includegraphics[width=13cm]{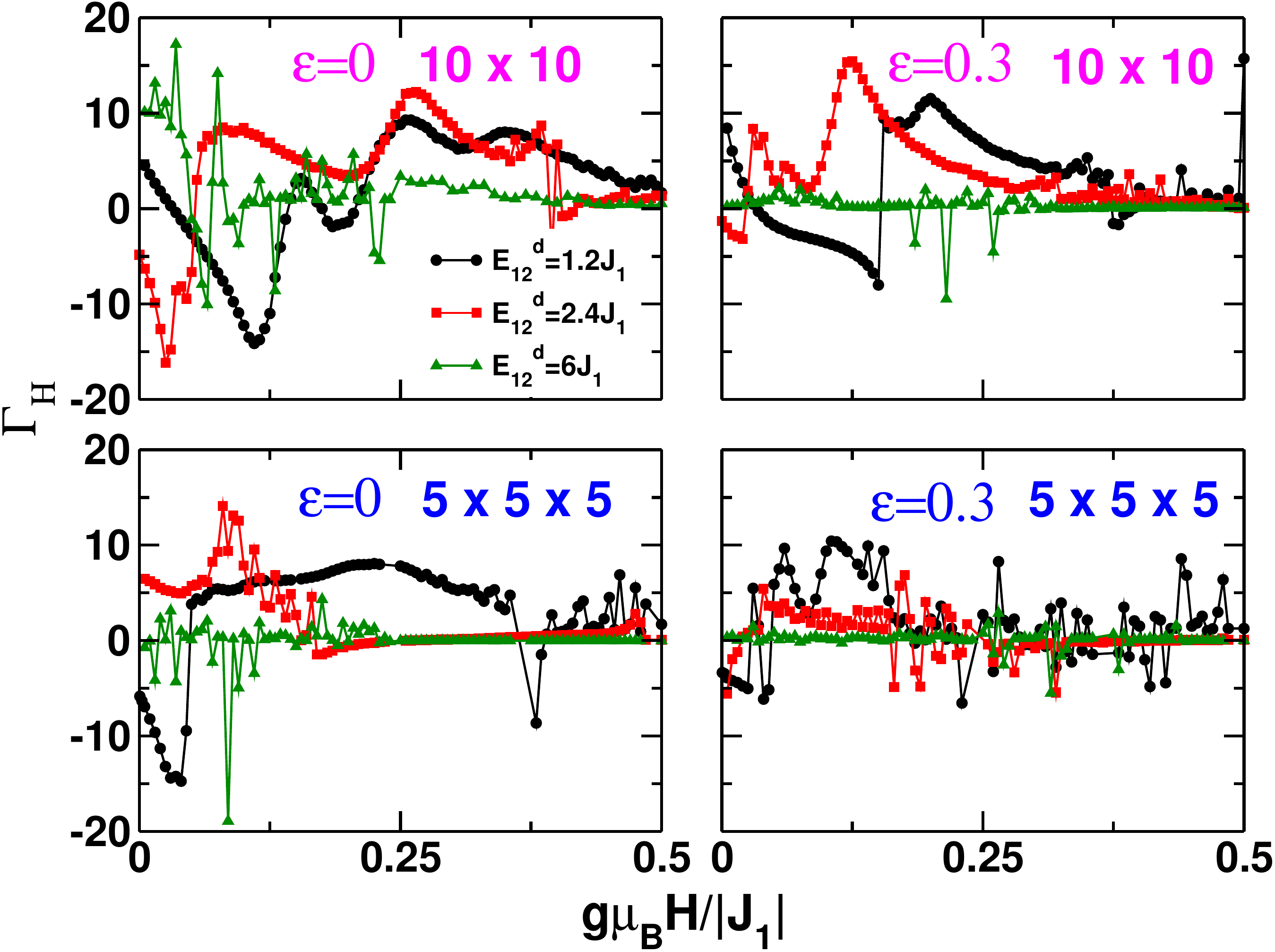}
      \caption{\label{fig:oneDtwoDGamma}Variation of Gr\"uneisen parameter, $\Gamma_H$, with applied magnetic field ($g\mu_B H/|J_1|$) for the two different systems with $N=100$ and $125$ on a square lattice and a cubic lattice respectively for different values of spin-dipolar interactions $E_{12}^d=1.2J_1$, $2.4J_1$ and $6J_1$ for exchange anisotropies $\epsilon=0$ and $0.3$ and on-site anisotropy $|d/J_1|=0.1$, for spin cluster with site spin $s=1$ at temperature $k_BT/|J_1|=0.1$.}
\end{figure}
In 2-d and 3-d systems (Fig. \ref{fig:oneDtwoDGamma}), increase in dipolar interaction strength in the isotropic model shifts singularities in $\Gamma_H$ to lower fields and in the strong anisotropic case ($\epsilon=0.3$), strong dipolar interaction suppresses the magnetocaloric effect almost completely, at low field. However we see small $\Gamma_H$ at low fields and sharp singularity at intermediate fields in both the cases for weak dipolar interactions.

\begin{figure}[hbt!]
    \includegraphics[width=8cm]{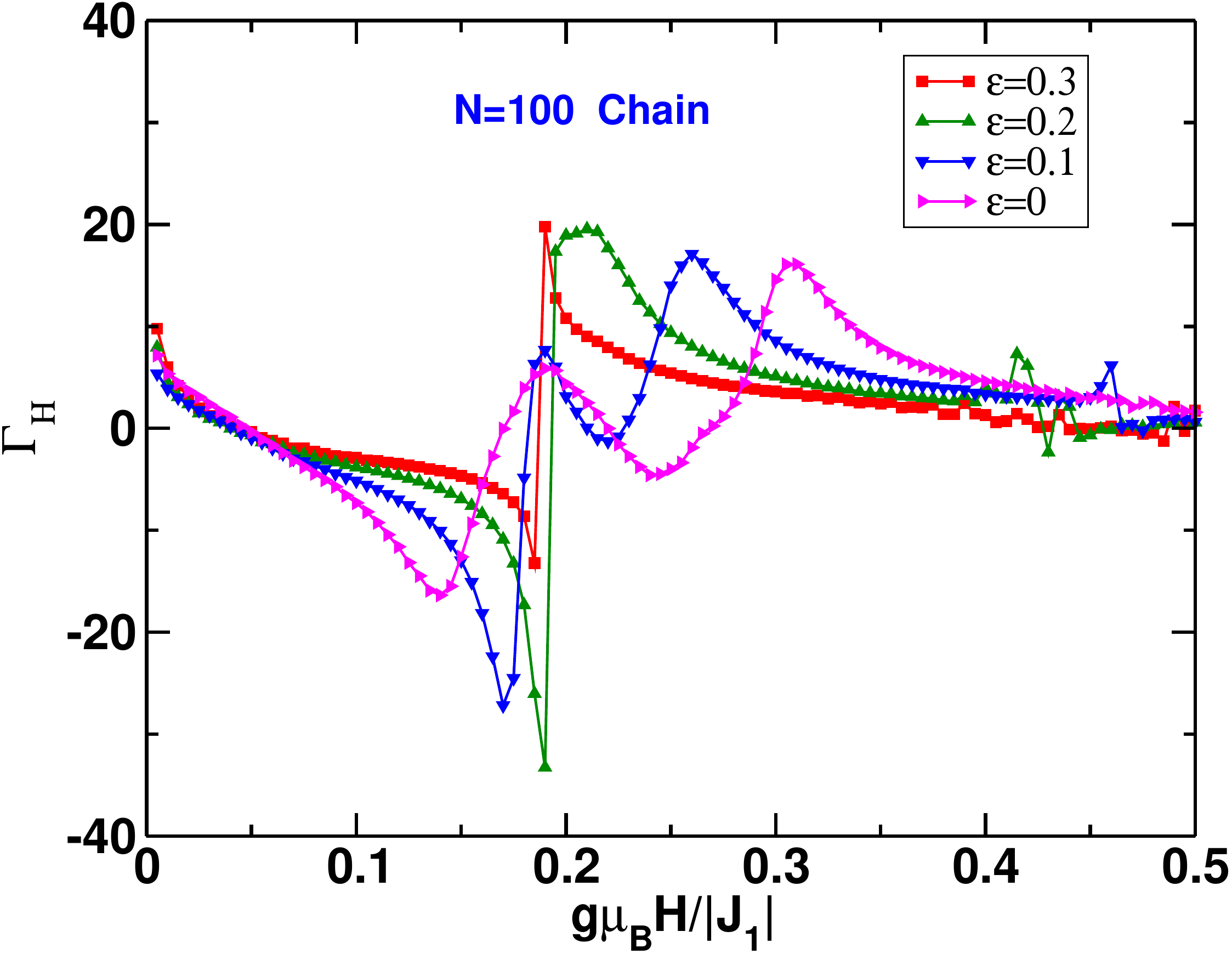} \\
     \includegraphics[width=8cm]{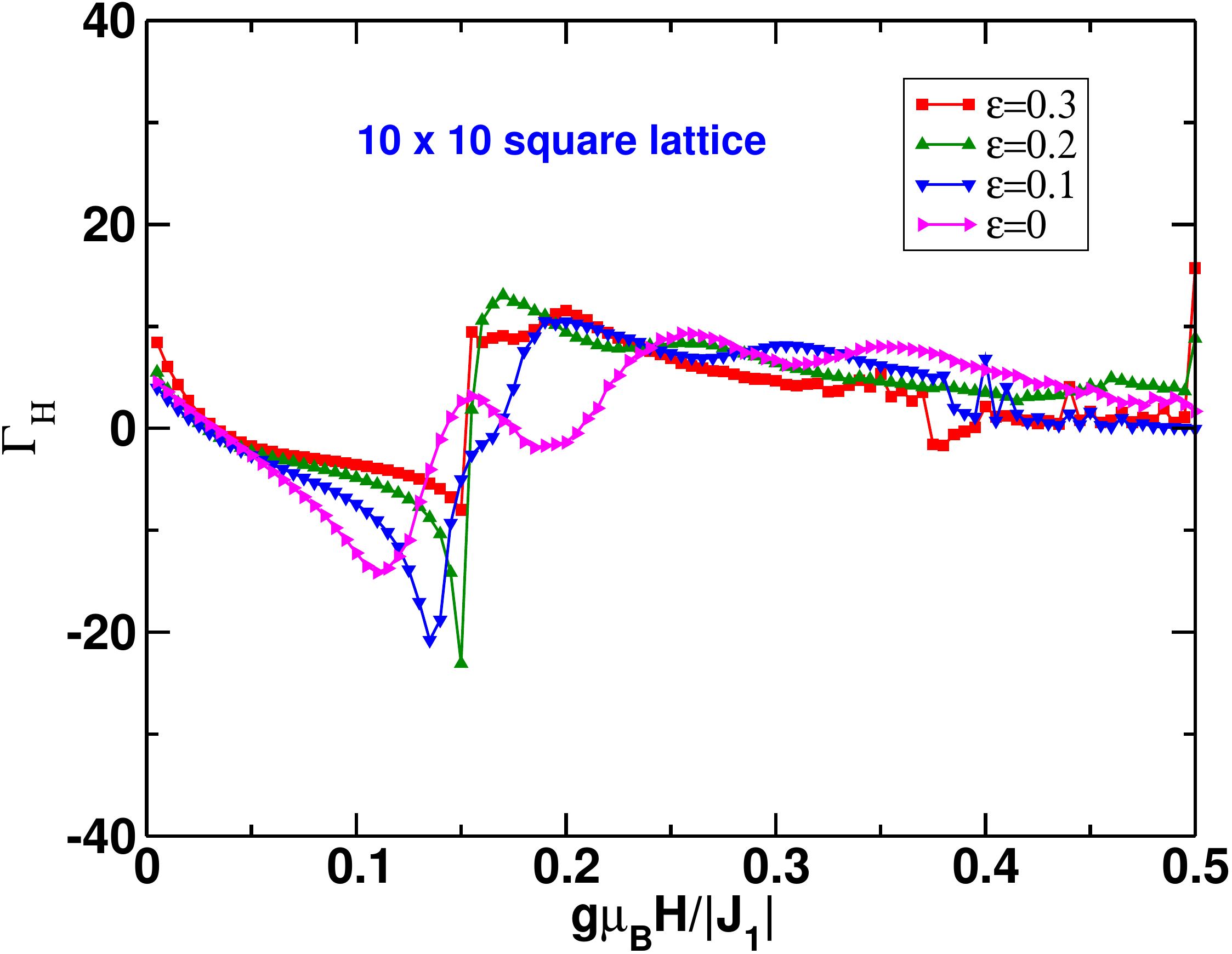}  \\
     \includegraphics[width=8cm]{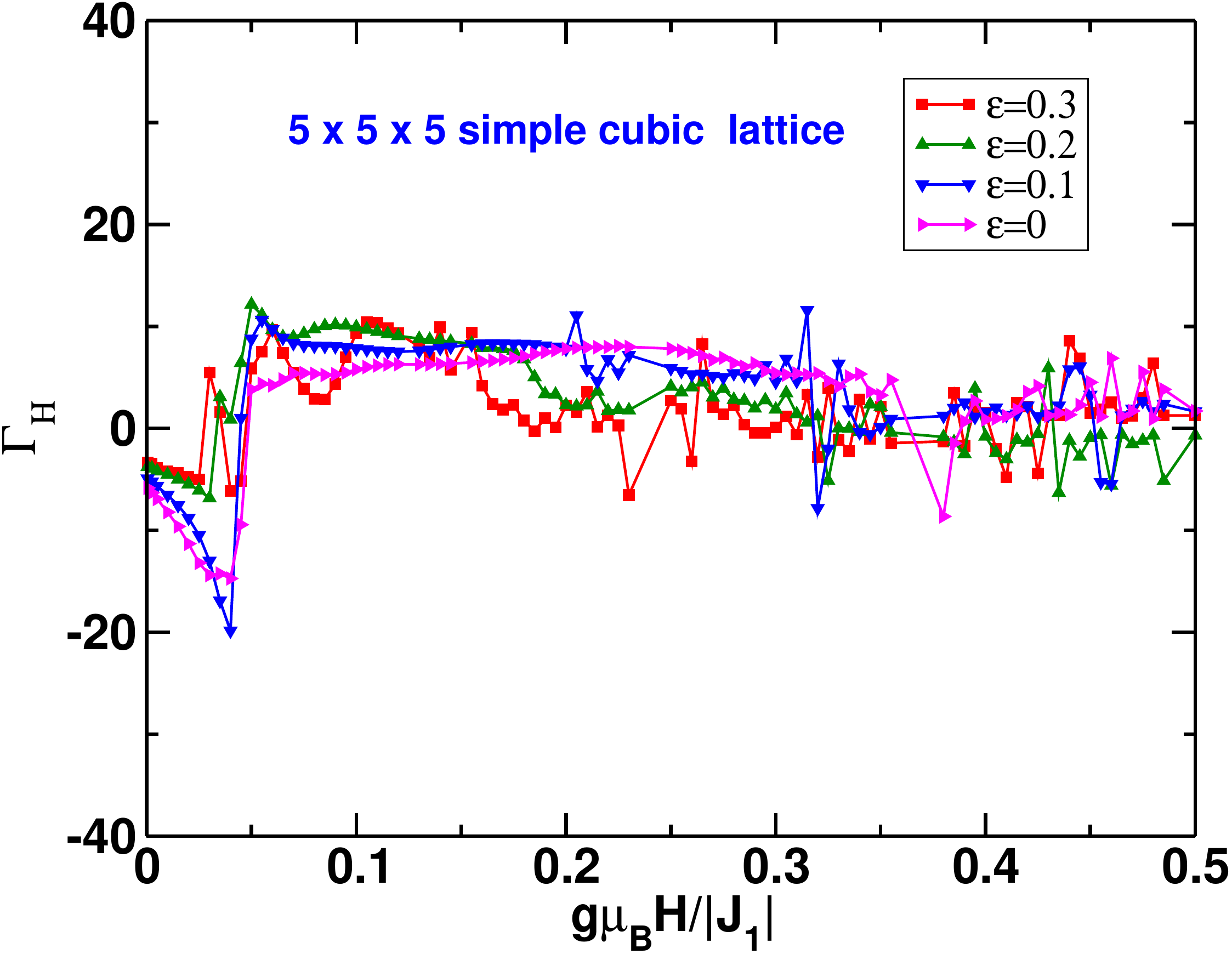} 
      \caption{\label{fig:MCGruneisen}Plots of Gr\"uneisen parameter, $\Gamma_H$, as a function of applied magnetic field ($g\mu_B H/|J_1|$) for an assembly of spin clusters on a chain with system size $N=100$, on $10 \times 10$ square lattice and $5 \times 5 \times 5$ simple cubic lattice for constant spin-dipolar interaction $E_{12}^d=1.2J_1$ for different exchange anisotropies $\epsilon$, for spin cluster with site spin $s=1$ at temperature $k_BT/|J_1|=0.1$.}
\end{figure}
In Fig. \ref{fig:MCGruneisen} we show the dependence of $\Gamma_H$ on dimensionality of the lattice for a fixed lattice constant corresponding to $E_{12}^d=1.2J_1$, for different anisotropies. In a chain, as the anisotropy of the exchange interaction increases, the singularity in $\Gamma_H$ shifts to higher fields and also becomes sharper. Besides, for the isotropic system there are two broad oscillations. The wavelengths of these oscillations decrease with increasing $\epsilon$ and for $\epsilon=0.2$ and $0.3$ there is only one sharp singularity but $\Gamma_H$ is smaller for $\epsilon=0.3$ than for $\epsilon=0.2$. When we go to a square lattice, while the general behaviour is similar to that of the chain, the singularities in $\Gamma_H$ for the same $\epsilon$ are at a lower field than for the corresponding chain. Besides $\Gamma_H$ in 2-d is smaller than in 1-d at the same field for corresponding systems. In 3-d the $\Gamma_H$ is the smallest and the first singularity is at a much lower field than in the 1-d and 2-d cases. Besides, the high field behaviour in 3-d shows many singularities. This can be understood from the fact that in 3-d a spin has more neighbors at the same distance than in 1-d or 2-d. This would mean many more micro states of spin orientations which are close in energy. Thus, a slight field shift can take the system from one spin configuration to another.

\subsection{\label{sec:Frustratedmodel}Frustrated spin clusters (Model II)}
\begin{figure}[hbt!]
    \includegraphics[width=8cm]{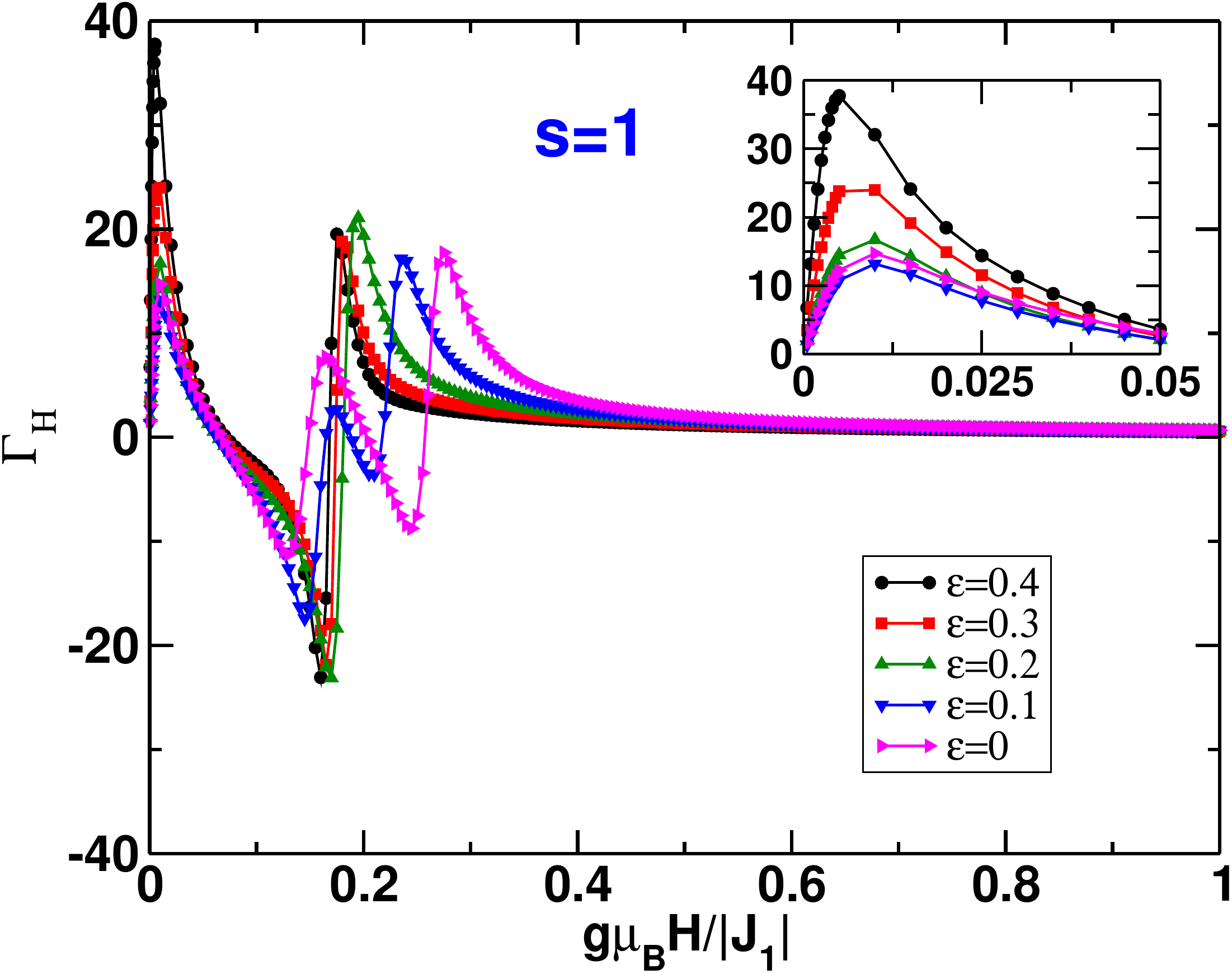} \\
     \includegraphics[width=8cm]{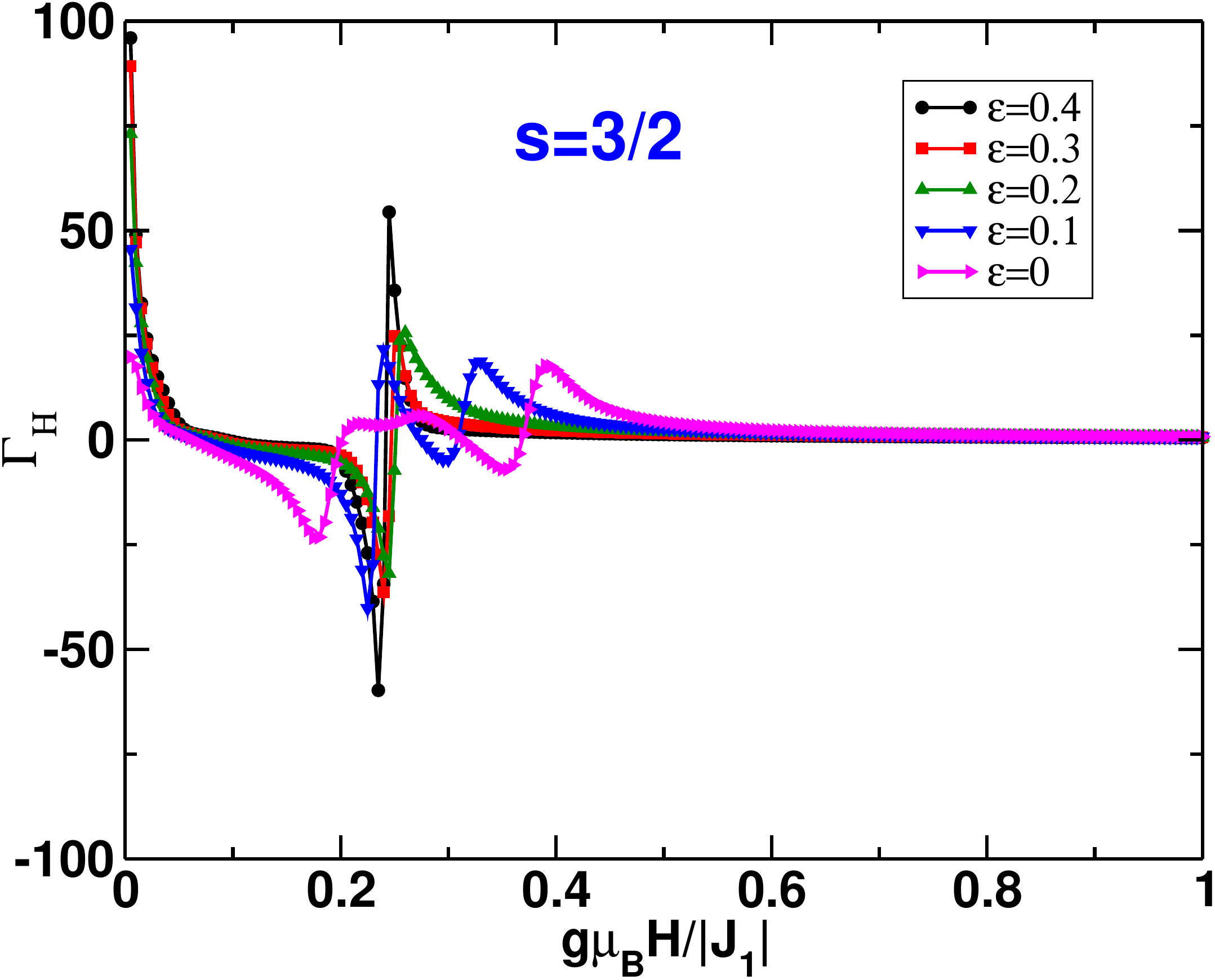}  \\
     \includegraphics[width=8cm]{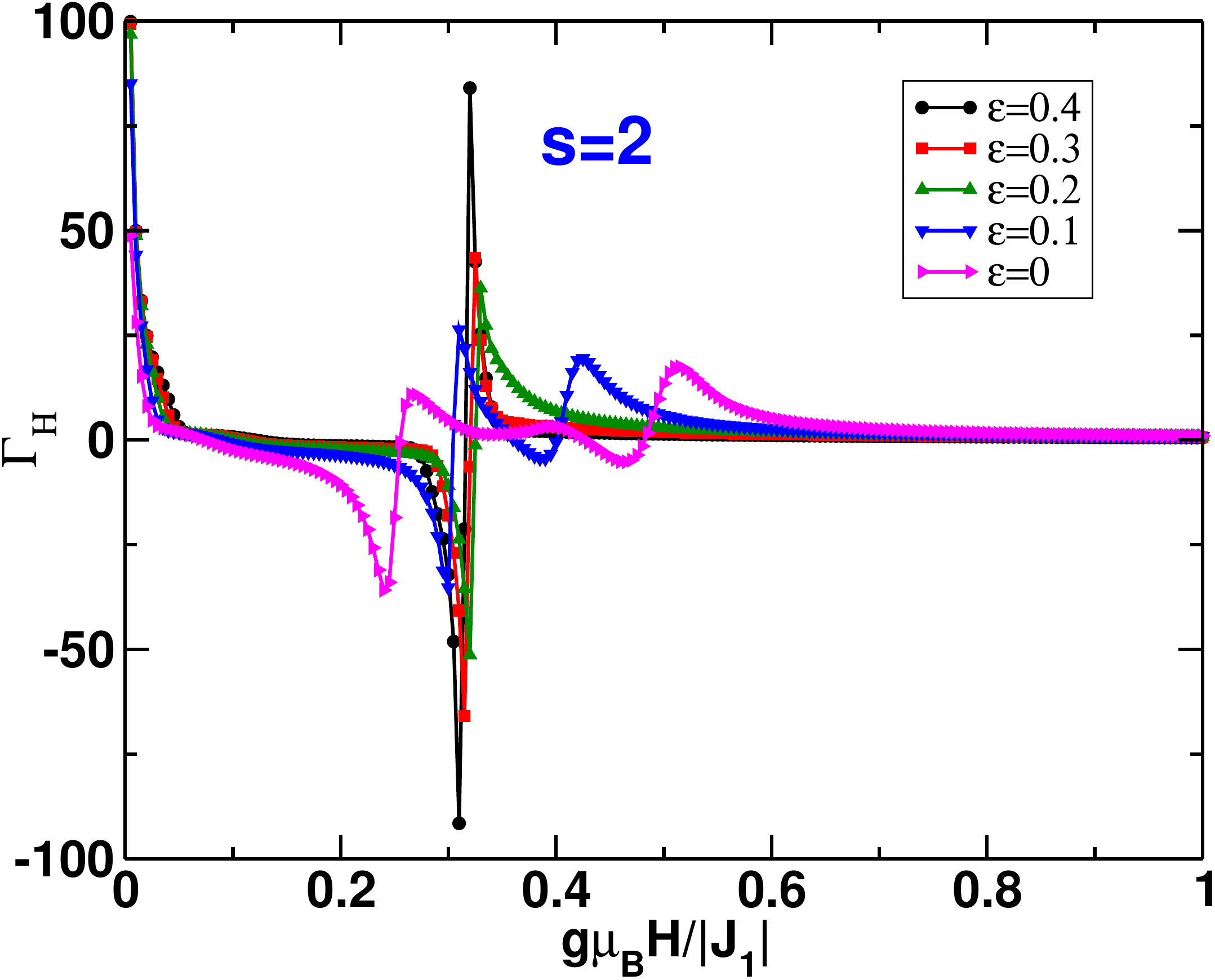} 
      \caption{\label{fig:FrustratedGruneisen}Plots of Gr\"uneisen parameter, $\Gamma_{H}$, as a function of applied magnetic field ($g\mu_B H/|J_1|$) in case of model II with next nearest neighbor isotropic ferromagnetic interactions ($J_{2}=0.2J_1$) at temperature $k_BT/|J_1|=0.1$ for different values of exchange anisotropy $\epsilon$, in the presence of on-site anisotropy $|d/J_1|=0.1$ for spin chains of six sites with site spins $s=1$, $3/2$, and $2$.}
\end{figure}

\begin{figure}[hbt!]
    \includegraphics[width=8cm]{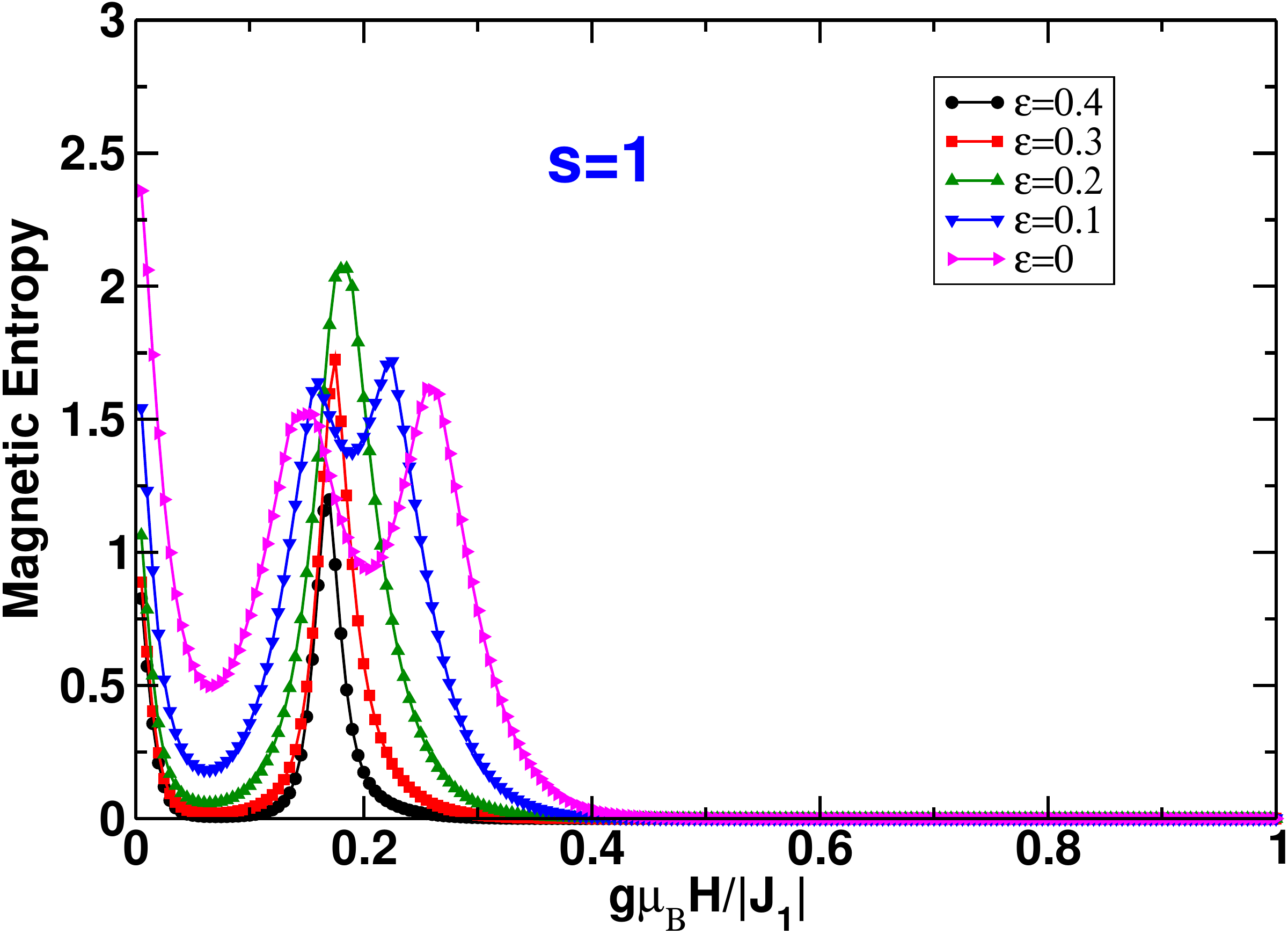} \\
     \includegraphics[width=8cm]{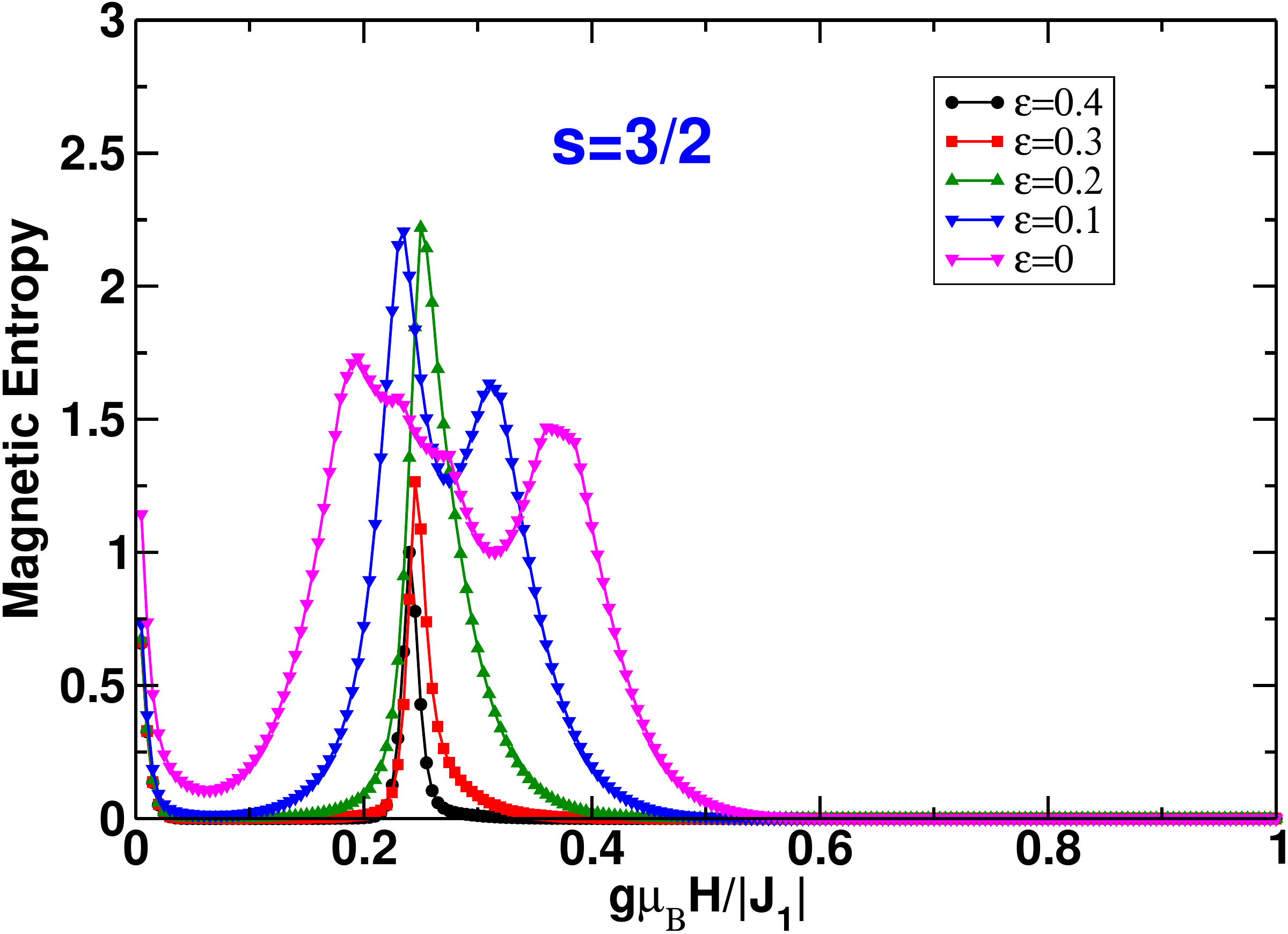} \\  
     \includegraphics[width=8cm]{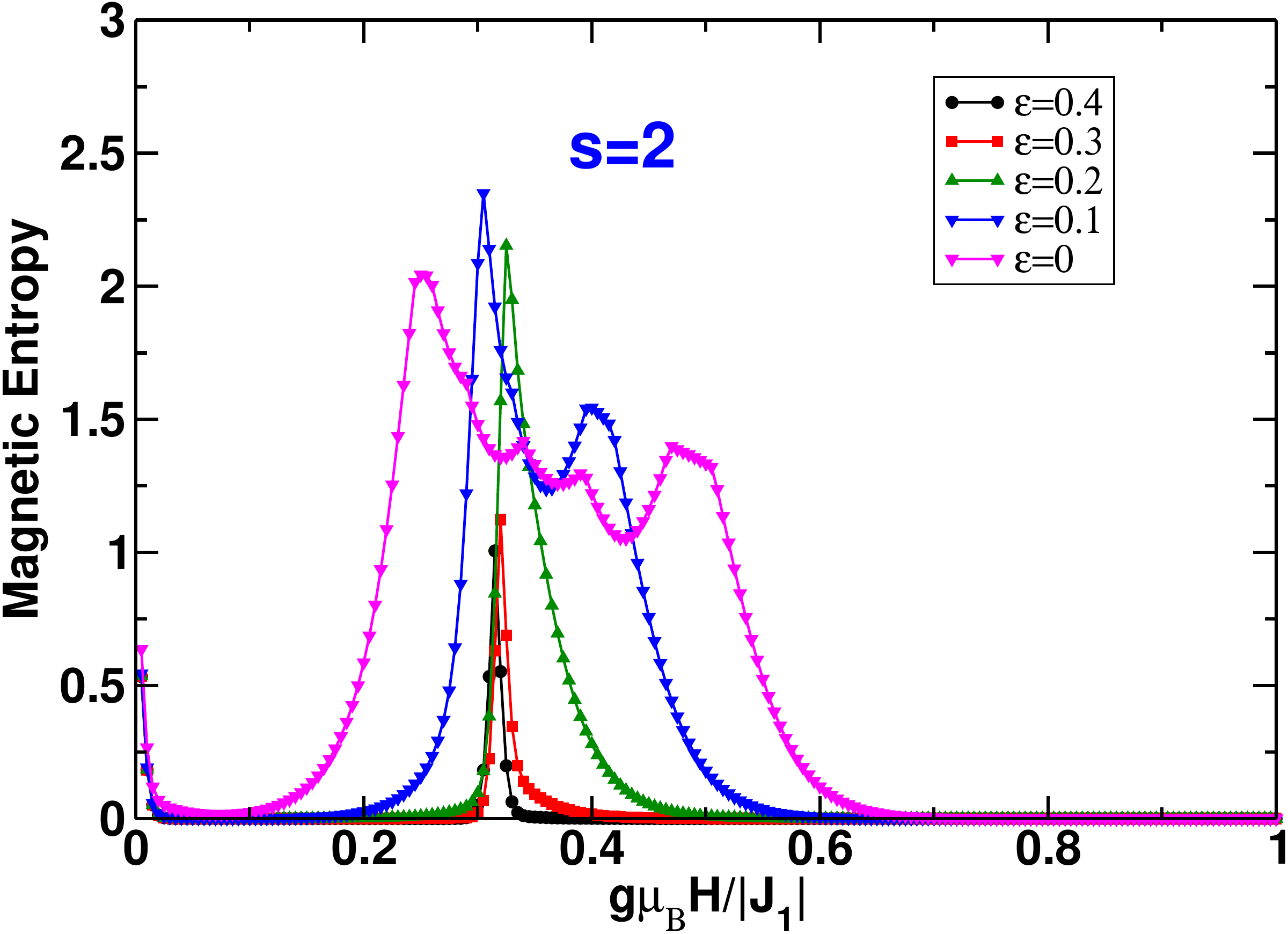} 
      \caption{\label{fig:Frustratedentropy}Variation of magnetic entropy, with applied magnetic field ($g\mu_B H/|J_1|$) at temperature $k_BT/|J_1|=0.1$ for different values of exchange anisotropy $\epsilon$, in the presence of on-site anisotropy $|d/J_1|=0.1$ for spin chains of six sites with site spins $s=1$, $3/2$, and $2$ for nnn isotropic ferromagnetic interactions ($J_{2}=0.2J_1$).}
\end{figure}

We have studied the magnetic Gr\"uneisen parameter in models with next nearest neighbor ferromagnetic interactions which result in spins frustration. In Fig. \ref{fig:FrustratedGruneisen}
we have shown the dependence of $\Gamma_H$ for a single spin cluster of site spins $s=1$, $3/2$ and $2$. In all the cases, we find large $\Gamma_H$ parameters in the low-field region with $\Gamma_H$ increasing with site spin as in model I. We also find that for the case of the isotropic model and  the weak anisotropic model ($\epsilon=0$, $0.1$), there are two broad oscillations in the $s=1$ and $3/2$ cases and three oscillations for $\epsilon=0$ and two oscillations for $\epsilon=0.1$ in the case of $s=2$ models. The magnetic fields at which these oscillations occur increases with increasing site spin. The magnitude of $\Gamma_H$ increases with increasing site spin in both the low field case and the intermediate field case. The magnetic entropy for the $s=1$ case shows two peaks, in the intermediate field region for $\epsilon=0$, $0.1$ and a single peak for larger $\epsilon$. In the $s=3/2$ and $2$ cases the isotropic model shows many cusps in the intermediate field regime. For higher anisotropies the entropy shows Fig. \ref{fig:Frustratedentropy} 
a sharp peak which is reflected as a sharp singularity in $\Gamma_H$ at the intermediate fields. In all cases there is an increase in the entropy near zero field. This behaviour is consistent with the observed $\Gamma_H$ dependence on the applied field. 

\begin{figure}[hbt!]
    \includegraphics[width=13cm]{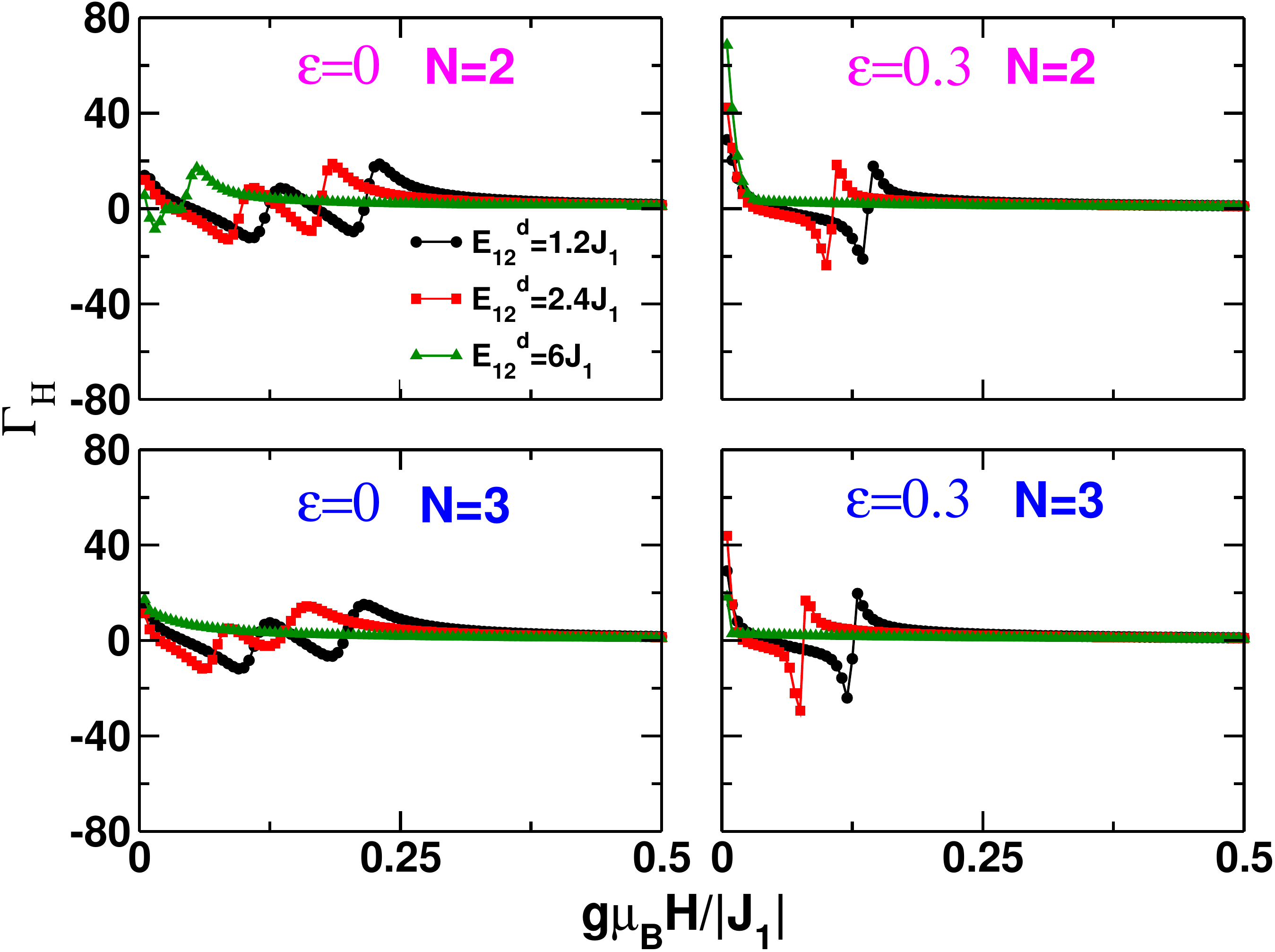}
      \caption{\label{fig:smallsystemGammaFrustd}Variation of Gr\"uneisen parameter, $\Gamma_H$, with applied magnetic field ($g\mu_B H/|J_1|$) for two small spin clusters $N=2$ and $3$ (on a chain), with different values of spin-dipolar interactions $E_{12}^d=1.2J_1$, $2.4J_1$ and $6J_1$ for exchange anisotropies $\epsilon=0$ and $0.3$ and on-site anisotropy $|d/J_1|=0.1$ with nnn isotropic ferromagnetic interactions ($J_{2}=0.2J_1$) for spin cluster with site spin $s=1$ at temperature, $k_BT/|J_1|=0.1$.}
\end{figure}

\begin{figure}[hbt!]
    \includegraphics[width=8cm]{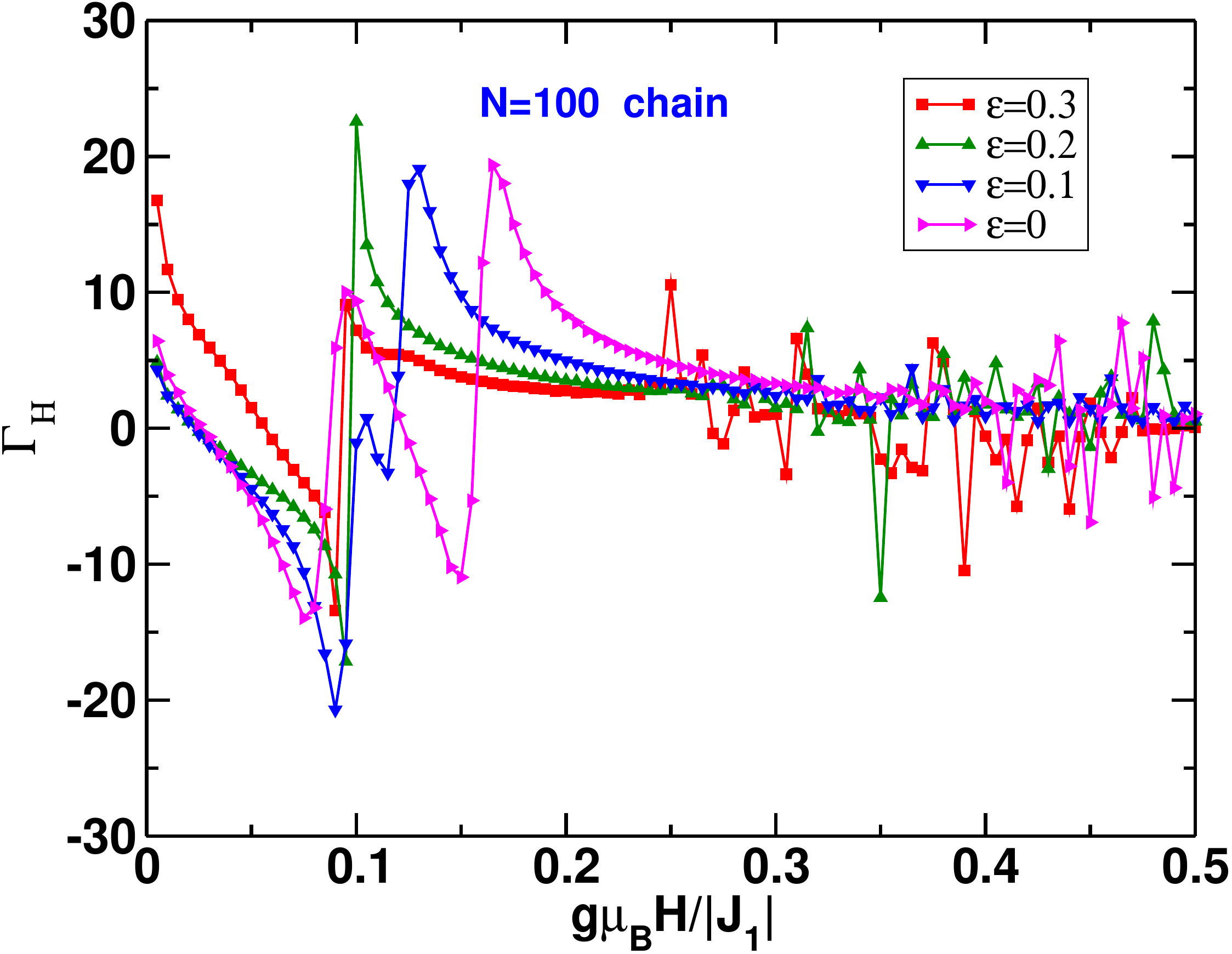} \\
     \includegraphics[width=8cm]{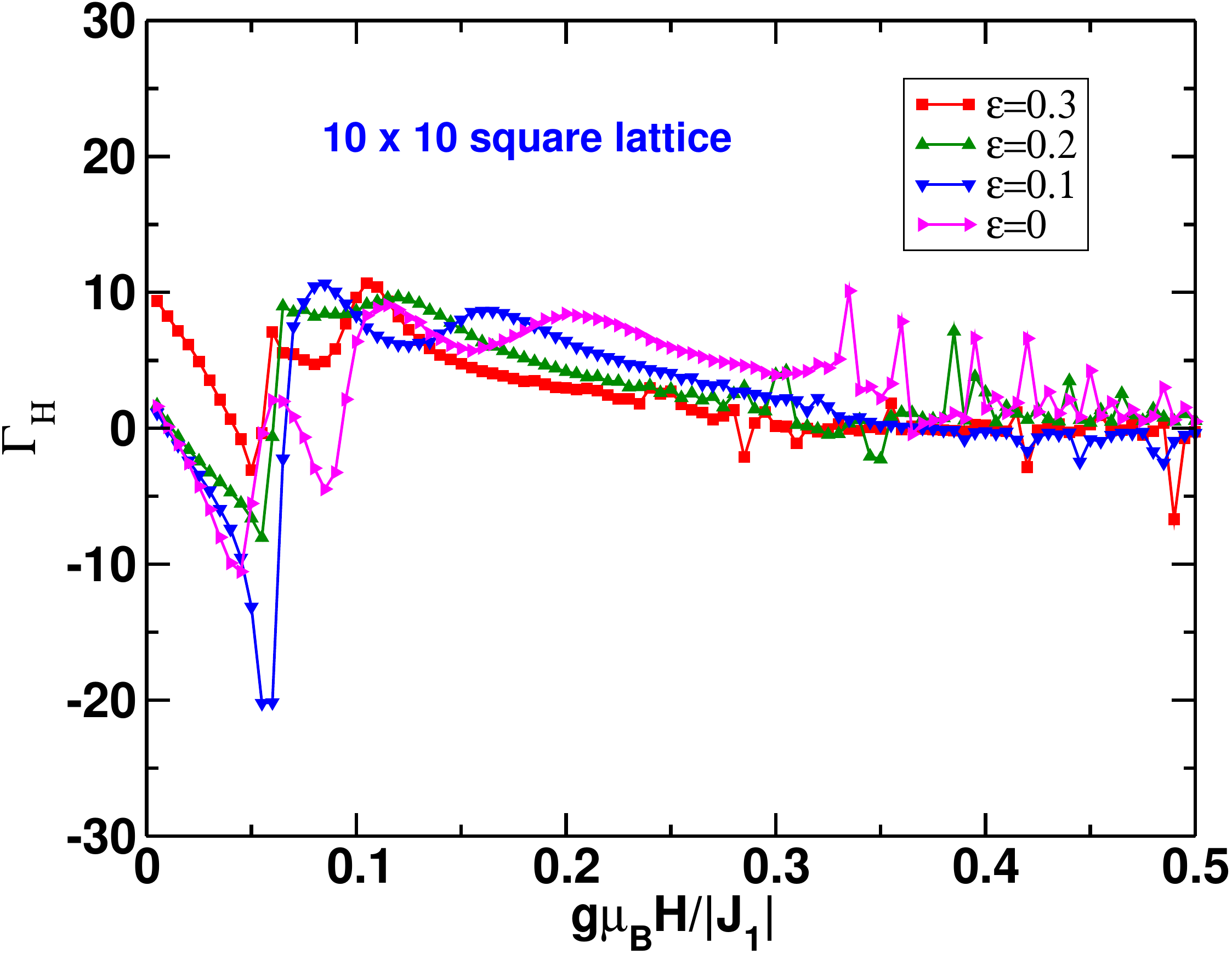}  \\
     \includegraphics[width=8cm]{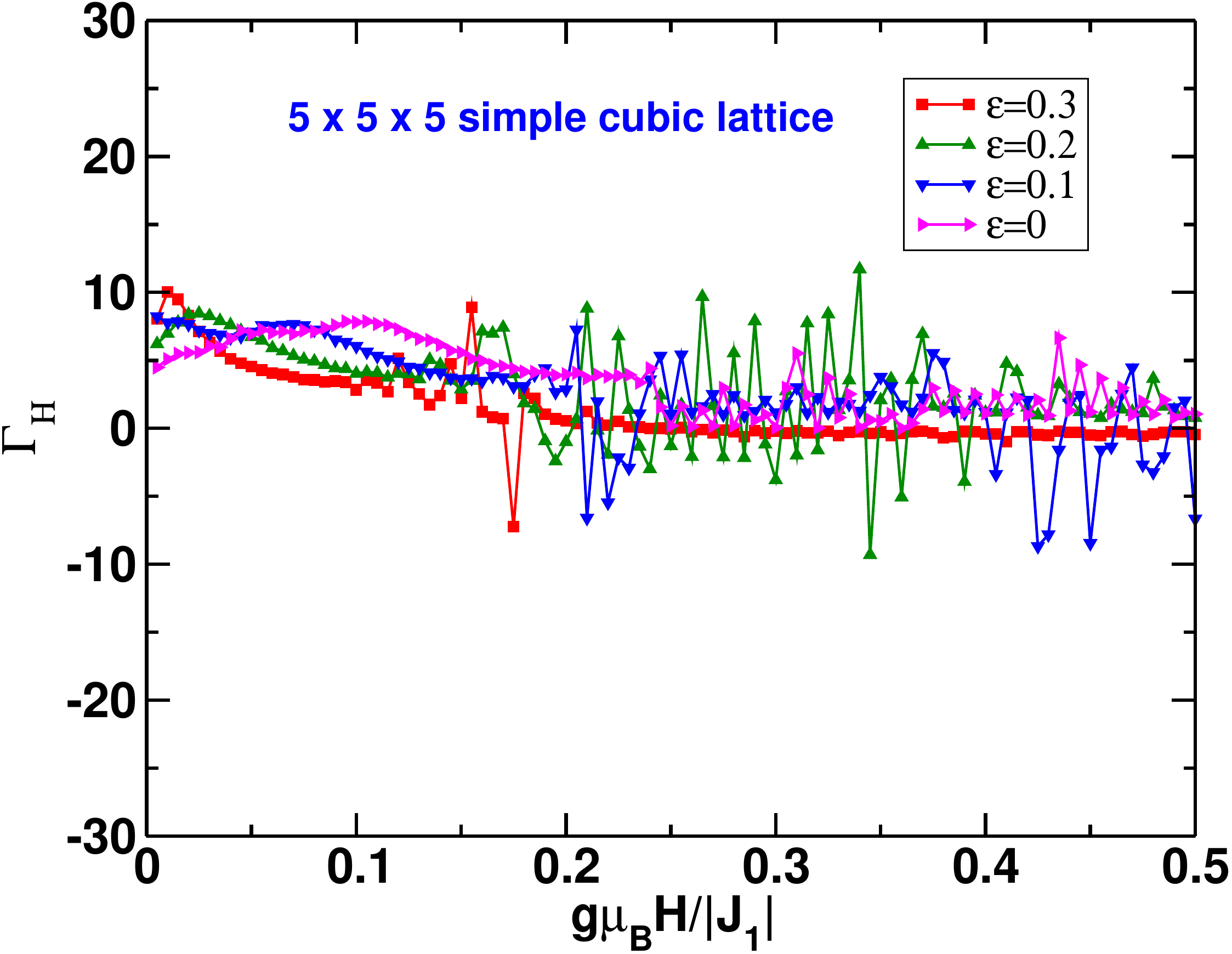} 
      \caption{\label{fig:MCGruneisenFrustd}Comparison of Gr\"uneisen parameter, $\Gamma_H$, for different dimensionalities. Applied magnetic field ($g\mu_B H/|J_1|$) for an assembly of spin clusters on a chain with system size $N=100$, on a $10 \times 10$ square lattice and a $5 \times 5 \times 5$ cubic lattice for constant spin-dipolar interaction $E_{12}^d=1.2J_1$ for different exchange anisotropies $\epsilon$, with nnn isotropic ferromagnetic interactions ($J_{2}=0.2J_1$) for spin cluster with site spin $s=1$ at temperature $k_BT/|J_1|=0.1$.}
\end{figure}

We have studied the spin-1 model in 1-D (upto 100 sites), in 2-D ($10 \times 10$ square lattice) and in 3-D ($5 \times 5\times 5$ simple cubic lattice). We have also studied the 1-D system for different cluster sizes. For a cluster of 2 and 3 molecules with isotropic exchange, the singularity in $\Gamma_H$ shift to lower fields as the interaction strength is increased (Fig. \ref{fig:smallsystemGammaFrustd}).
The number of oscillations in intermediate fields decrease with increase in inter-molecular spin dipolar interactions. When the exchange anisotropy is strong, the oscillations in $\Gamma_H$ become sharp singularities and shift to lower fields as the inter cluster interactions become stronger. For larger clusters ($N=10$ and $100$)
\begin{figure}[hbt!]
    \includegraphics[width=13cm]{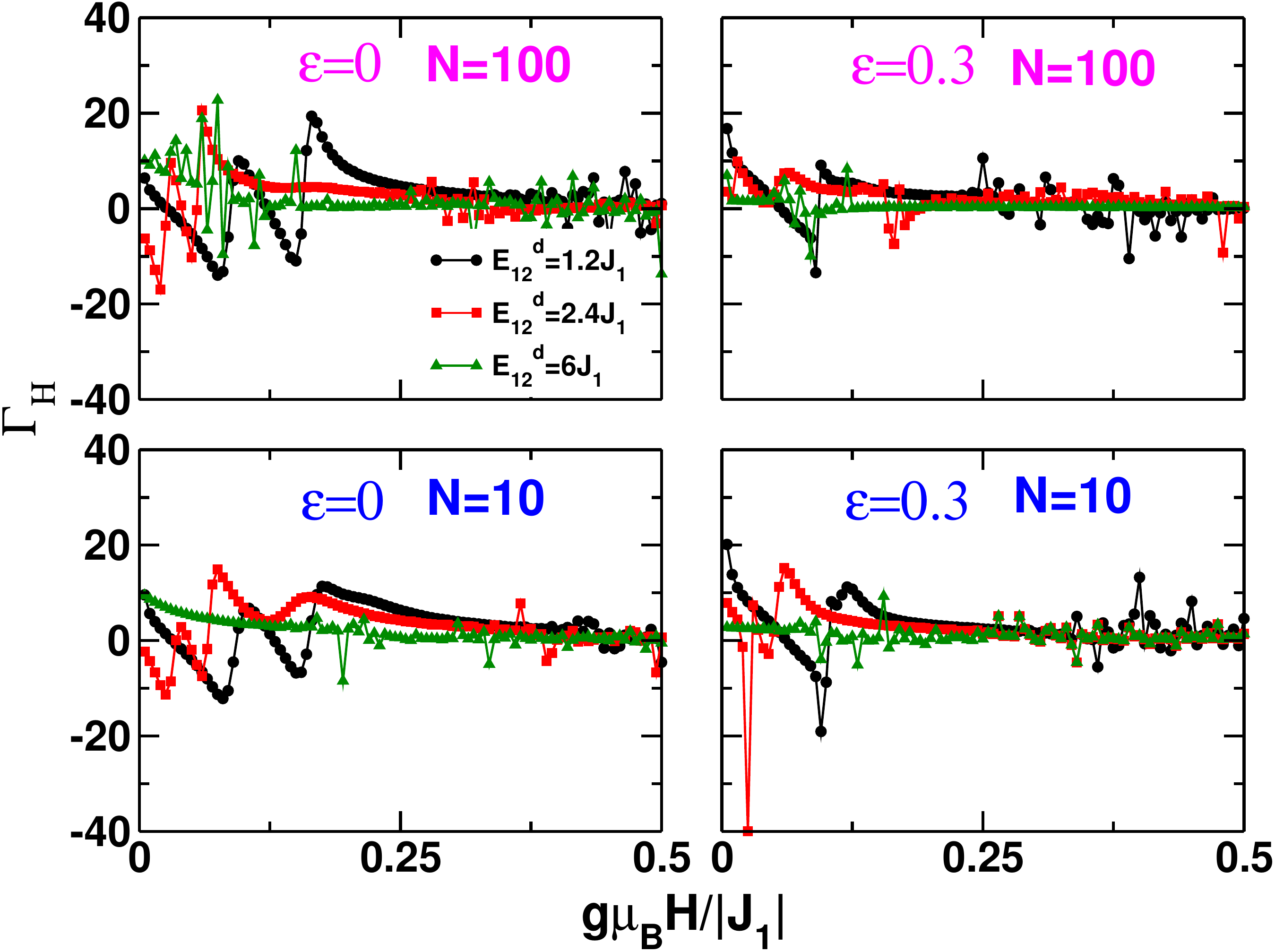}
      \caption{\label{fig:largesystemGammaFrustd}Variation of Gr\"uneisen parameter, $\Gamma_H$, with applied magnetic field ($g\mu_B H/|J_1|$) for two spin clusters with $N=100$ and $10$ on a chain with different values of spin-dipolar interactions $E_{12}^d=1.2J_1$, $2.4J_1$ and $6J_1$ for exchange anisotropies $\epsilon=0$ and $0.3$ and on-site anisotropy $|d/J_1|=0.1$ with nnn isotropic ferromagnetic interactions ($J_{2}=0.2J_1$) for spin cluster with site spin $s=1$ at temperature $k_BT/|J_1|=0.1$.}
\end{figure}
 increasing the inter-molecular interactions strength shift the oscillations to lower fields and at higher fields, in the $N=100$ clusters many closely lying singularities are found (Fig. \ref{fig:largesystemGammaFrustd}). Again, these singularities become sharper for large exchange anisotropy and appear to be much weaker for strong inter-molecular interactions. We have compared the $\Gamma_H$ behaviour as a function of the anisotropy in 1-D, 2-D and 3-D systems in Fig. \ref{fig:MCGruneisenFrustd}.
In 1-D, at very low-field strength, the $\Gamma_H$ is larger in 1-D than in 2-D and the 2-D $\Gamma_H$ is larger than the $\Gamma_H$ in 3-D. In 1-D for all values of the anisotropy in exchange, there is a singularity at $g\mu_BH/J_1 \sim 0.08$ to $0.09$ and a subsequent singularities are $g\mu_BH/J_1 \sim 0.14$ to $0.15$. In 2-D these singularities shift to $\sim 0.06$. In 3-D there is a broad hump in $\Gamma_H$ whose maxima shifts from $\sim 0.12$ for $\epsilon=0$ to $0.02$ as $\epsilon$ is increased to $0.03$. Systems with exchange anisotropy, $\epsilon=0.1$ and $0.2$, show many oscillation in $\Gamma_H$ which are smoothened as $\epsilon$ increases to $0.3$. For the isotropic models, in 2-D the oscillation in $\Gamma_H$ become sharper and shift to higher fields as the inter-molecular interaction strength increases. While in 3-D, from a broad oscillation the $\Gamma_H$ shows sharp variation at intermediate inter-molecular interactions, which eventually disappears for large inter-molecular interaction strengths. A similar behaviour is also found when the exchange anisotropy is large.

\section{\label{sec:conclusion}Conclusions}
\vspace*{-0.5cm}
We have calculated exact magnetic Gr\"uneisen parameters $\Gamma_H$ for two different spin models (i) spin chains with alternating ferro and antiferro magnetic exchange interactions and (ii) frustrated spin chains with additional next nearest neighbor ferromagnetic interaction. For efficient cooling by AD, we need systems with large $\Gamma_H$. In isolated spin chains, we have seen two different characteristics in high and low applied magnetic field regimes. $\Gamma_H$ exhibits peaks in low field for all spin chains and these peaks are higher for larger exchange anisotropy. In high field region, we have observed oscillations in $\Gamma_H$ and these oscillations become singularities with increase of exchange anisotropy and these singularities shift to higher fields as site spin increases. The singularities in $\Gamma_H$ correspond to cusps in the magnetic entropy plot in all cases and also coincide with crossovers in the ground state magnetization in the Zeeman plots. Furthermore, we have systematically studied the behaviour of magnetic Gr\"uneisen parameters in an assembly of spin clusters on a chain of 100 clusters, a $10 \times 10$ square lattice and a $5 \times 5 \times 5$ cubic lattice for spin-1 systems with inter-chain spin-dipolar interactions using Monte Carlo method. For all system sizes ($N=2$, $3$, $10$ and $100$), the singularities in $\Gamma_H$ shift to lower fields with increase in spin-dipolar interaction strength and in case of large system the singularity vanishes for $E_{12}^{dip}=6J_1$. These singularities become sharper with the increase in exchange anisotropy. We have also studied the dependence of $\Gamma_H$ on the dimensionality of the lattice for a fixed spin-dipolar interaction strength, $E_{12}^d=1.2J_1$. We observe that as the exchange anisotropy increases, the singularities in $\Gamma_H$ shift to higher fields in chains while in square and cubic lattice, these singularities shift to lower fields although $|\Gamma_H|$ becomes smaller. Similar behaviour is seen in frustrated spin model, although the magnitude of $\Gamma_H$ is smaller than in the corresponding non-frustrated spin model. Because of built in magnetic entropy in the frustrated model, the net change in magnetic entropy due to adiabatic demagnetization will be less sharp implying lower cooling efficiency in frustrated systems.
\section{\label{sec:ack}Acknowledgements}
SR thanks INSA for a senior scientist position and DST India for support through project EMR/2016/005183.

\end{document}